\documentclass[11pt,letterpaper]{article}

\usepackage[includeheadfoot,
            marginratio={1:1,2:3}, 
            width=412pt,
            height=688pt,]{geometry}
            \usepackage{tikz}
\usepackage[compat=1.1.0]{tikz-feynman}
\usepackage[utf8]{inputenc}
\usepackage{amsmath}
\usepackage{amsfonts}
\usepackage{amssymb}
\usepackage{graphicx}
\usepackage{booktabs}
\usepackage{empheq}
\usepackage{paralist}
\usepackage{cite}
\usepackage[hidelinks]{hyperref}
\usepackage{xcolor}
\usepackage{tensor}
\usepackage{adjustbox}
\usepackage{subcaption}
\usepackage{braket}
\usepackage{tikz}
\usepackage{tikz-cd}
\usepackage{yfonts}
\usepackage{enumitem}
\usepackage{physics}
\usepackage{slashed}
\usepackage{mdframed}

\usetikzlibrary{decorations.pathreplacing,calc}

\newcommand{\nc}{\newcommand}
\nc{\lb}{\llbracket}
\nc{\rb}{\rrbracket}
\nc{\gl}{\llbracket}
\nc{\gr}{\rrbracket}
\nc{\bbR}{\mathbb{R}}
\nc{\bbC}{\mathbb{C}}
\nc{\bbZ}{\mathbb{Z}}
\nc{\cO}{\mathcal{O}}
\nc{\cS}{\mathcal{S}}
\nc{\cM}{\mathcal{M}}
\nc{\cT}{\mathcal{T}}
\nc{\cX}{\mathcal{X}}
\nc{\cQ}{\mathcal{Q}}
\nc{\cA}{\mathcal{A}}
\nc{\cD}{\mathcal{D}}
\nc{\cL}{\mathcal{L}}
\nc{\cC}{\mathcal{C}}
\nc{\cG}{\mathcal{G}}
\nc{\cF}{\mathcal{F}}
\nc{\cI}{\mathcal{I}}
\nc{\cN}{\mathcal{N}}
\nc{\pd}{\partial}
\nc{\la}{\lambda}
\nc{\cU}{\mathcal{U}}

\newcommand\beq{\begin{equation}}
\newcommand\eeq{\end{equation}}
\nc{\del}{\partial}
\nc{\tri}{\hspace{-3.5pt}\vartriangle\hspace{-3.5pt}}
\nc{\blacktri}{\blacktriangle}
\nc{\eq}[1]{\begin{equation}
                     \begin{split} #1 \end{split}
                     \end{equation}}
\nc{\ul}{\underline}
\nc{\ov}{\overline}
\nc{\fa}{\hat}
\nc{\fb}{\MakeUppercase}
\nc{\fc}{\tilde }
\nc{\Lie}{{\cal L}} 
\nc{\lambdabar}{{\mkern0.75mu\mathchar '26\mkern -9.75mu\lambda}}

\newmuskip\pFqmuskip

\newcommand*\pFq[7][8]{
  \begingroup 
  \pFqmuskip=#1mu\relax
  \mathchardef\normalcomma=\mathcode`,
  \mathcode`\,=\string"8000
  \begingroup\lccode`\~=`\,
  \lowercase{\endgroup\let~}\pFqcomma
  {}_{#2}{#3}_{#4}{\left[\left.\genfrac..{0pt}{}{#5}{#6}\right|#7\right]}
  \endgroup
}
\newcommand{\pFqcomma}{{\normalcomma}\mskip\pFqmuskip}

\allowdisplaybreaks[2]

\nc{\Unote}[1]{\textcolor{gray}{#1}}

\begin{document}

\vspace*{1.5cm}

\begin{center}
{\LARGE
\textbf{Tame Complexity of Effective Field Theories \\
\vspace{0.25cm} 
in the Quantum Gravity Landscape}}

\end{center}

\vspace{0.35cm}
\begin{center}
 {\bf Thomas W.~Grimm$^{1,2,3}$},
 {\bf David Prieto$^{1,2}$},
 {\bf Mick van Vliet$^{1,2}$}
\end{center}

\vspace{.5cm}
\begin{center} 
\vspace{0.25cm} 
\emph{
$^1$\,
Institute for Theoretical Physics, Utrecht University,
\\
Princetonplein 5, 3584 CC Utrecht, 
The Netherlands } \\
\vspace{0.25cm} 
\emph{
$^2$\,
Center of Mathematical Sciences and Applications,\\ 
Harvard University, Cambridge, MA 02138, USA } \\
\vspace{0.25cm} 

$^3$\,
\emph{Jefferson Physical Laboratory,\\
Harvard University, Cambridge, MA 02138, USA}

\end{center}

\vspace{2.5cm}


\begin{abstract}
\noindent
Effective field theories consistent with quantum gravity obey surprising finiteness constraints, appearing in several distinct but interconnected forms. In this work we develop a framework that unifies these observations by proposing that the defining data of such theories, as well as the landscape of effective field theories that are valid at least up to a fixed cutoff, admit descriptions with a uniform bound on complexity. To make this precise, we use tame geometry and work in sharply o-minimal structures, in which tame sets and functions come with two integer parameters that quantify their information content; we call this pair their tame complexity. 
Our Finite Complexity Conjectures are supported by controlled examples in which an infinite Wilsonian expansion nevertheless admits an equivalent finite-complexity description, typically through hidden rigidity conditions such as differential or recursion relations. We further assemble evidence from string compactifications, highlighting the constraining role of moduli space geometry and the importance of dualities. This perspective also yields mathematically well-defined notions of counting and volume measures on the space of effective theories, formulated in terms of effective field theory domains and coverings, whose finiteness is naturally enforced by the conjectures.

\end{abstract}

\clearpage

\tableofcontents


\newpage

\parskip=.2cm

\section{\label{sec:intro}Introduction}

A unifying theme running through quantum gravity constraints on effective theories is \textit{finiteness}. In various forms, it plays a central role in the swampland program, which aims to characterize which low-energy effective field theories (EFTs) admit a consistent UV completion including quantum gravity \cite{Vafa:2005ui} (we refer to \cite{Palti:2019pca,vanBeest:2021lhn,Agmon:2022thq} for reviews).
From this viewpoint, the space of candidate EFTs splits into a landscape of theories that do arise from quantum gravity and a swampland of apparently consistent EFTs that nonetheless cannot be completed into quantum gravity. The principle of finiteness has various realizations in this context, and constrains the size of the landscape \cite{Vafa:2005ui,Acharya:2006zw}, the matter spectrum \cite{Kumar:2010ru,Lee:2019skh,Kim:2019vuc,Katz:2020ewz,Tarazi:2021duw,Martucci:2022krl,Kim:2024eoa,Birkar:2025rcg}, geometric features of moduli spaces \cite{Ooguri:2006in,McNamara:2019rup}, and amplitudes \cite{Hamada:2021yxy}. It is therefore natural to seek a unifying notion that relates these finiteness constraints and clarifies their common origin in quantum gravity and its imprints on low-energy physics. Taken together, these interconnected finiteness phenomena suggest that EFTs in the landscape are not arbitrarily complicated, but instead have a bounded complexity. Yet, at present, there is no precise definition of EFT complexity. In this work we propose a framework to make this intuition quantitative by introducing a notion of complexity for EFTs and formulating a conjecture that EFTs in the quantum gravity landscape have finite complexity.

As a first step toward such a framework, we seek a notion of finiteness that is mathematically precise and robust under the standard operations that relate EFT descriptions (e.g.~integrating out degrees of freedom, renormalization group flow, and field redefinitions).
A natural starting point is using the concept of tameness, which was proposed in \cite{Grimm:2021vpn} as a universal finiteness principle that can unify many of the finiteness phenomena appearing in the quantum gravity landscape. Conceptually, tameness restricts attention to sets and functions with controlled geometric behavior, excluding, for instance, arbitrarily wild oscillations or infinitely intricate boundaries, and thereby provides a setting in which finiteness properties are preserved under the operations relevant in physics.
Mathematically, tameness is formalized by the theory of o-minimal structures, originating in mathematical logic, and it first found physical applications through finiteness questions in F-theory flux vacua \cite{Grimm:2021vpn,Bakker:2023xkt}. More recently, it was shown in \cite{Douglas:2022ynw,Douglas:2023fcg,Grimm:2024hdx} that many quantities in quantum field theory (QFT), including perturbative scattering amplitudes, naturally fit into this tame framework.

To go beyond qualitative finiteness and obtain a quantitative notion of complexity, we use a refined version of tameness known as sharp o-minimality \cite{binyamini2022sharply,beyondo-min}. The striking feature of this refinement is that it assigns to each tame set or function a pair of integers $(\cF,\cD)$ -- called format and degree -- which, in a chosen representation, quantify the amount of information needed to specify it. We refer to this pair as the \textit{tame complexity}, to distinguish it from other notions of complexity (reviewed, e.g.~in~\cite{Baiguera:2025dkc}). Importantly, these integers are designed to behave predictably under the basic operations by which more complicated objects are built from simpler ones, 
so that complexity bounds propagate to derived constructions. In this way, tame complexity provides a robust quantitative handle on geometric and computational features of tame objects, in a spirit analogous to the role of degree in algebraic geometry. These ideas were recently applied to quantum field theories in \cite{Grimm:2023xqy,Grimm:2024mbw,Grimm:2024elq}. In particular, \cite{Grimm:2024elq} proposed to characterize the complexity of a QFT by studying the tame complexity of basic defining data, such as its Lagrangian  or its scattering amplitudes.

The aim of this paper is to use these recent advances to ask whether tame complexity can be used to distinguish EFTs in the quantum gravity landscape and the swampland. 
We argue that it can, and we formulate a \textbf{Finite Complexity Conjecture} in two steps: a \emph{local} version asserting that each EFT consistent with quantum gravity admits a description of finite tame complexity $(\cF_{\rm EFT},\cD_{\rm EFT})$, and a stronger \emph{global} version asserting that for EFTs valid at least up to a fixed cutoff $\Lambda$ these complexities are uniformly bounded across the landscape by $(\mathfrak{F}_\Lambda,\mathfrak{D}_\Lambda)$. Moreover, the global conjecture asserts that the set of such EFTs  
itself has finite $\Lambda$-dependent tame complexity.
By contrast, an unconstrained Wilsonian EFT typically requires specifying infinitely many independent couplings (Wilson coefficients), and therefore need not admit any description of finite tame complexity. 
More generally, the conjecture sharply constrains would-be discrete infinities in the landscape: if, at fixed $\Lambda$, one had infinitely many inequivalent EFTs or a single EFT interpolating among infinitely many distinct vacua, this would force the relevant tame complexity to diverge.
Taken together, these statements provide a unified quantitative perspective on finiteness constraints in the landscape, and a framework for analyzing their interrelations in more detail.

To formulate these conjectures, we must first extend the discussion of \cite{Grimm:2024elq} and explain how to assign tame complexity to general EFTs. This is a non-trivial step, since a Wilsonian EFT is typically specified by an infinite tower of higher-dimension operators and couplings and hence has, \emph{a priori}, infinite tame complexity. Based on several exactly solvable examples, we will show that in special cases the EFT data can nevertheless be reorganized into an equivalent description of finite tame complexity. A common mechanism behind this reduction is the presence of \emph{finite} differential constraints: if the  EFT couplings are governed by certain finite systems of differential equations with finitely many parameters, as happens in many EFTs arising in string compactifications or which have a geometric representation, then infinitely many Wilson coefficients become related and can be encoded implicitly in a finite amount of data. Building on swampland considerations, we then argue that EFTs consistent with quantum gravity fall into such special classes, so that a finite-complexity description exists despite the presence of infinitely many Wilsonian couplings.

Another essential step is to account for parameter dependence and the geometry of moduli spaces. As one moves in moduli space, the effective description can change qualitatively and eventually cease to be valid near its boundaries \cite{Ooguri:2006in}. In particular, there is generally no single EFT action that remains valid globally. We therefore introduce the notion of \emph{EFT domains} and \emph{EFT coverings}: a collection of regions in parameter space, each equipped with a local EFT description of finite tame complexity, whose union covers the relevant moduli/parameter space. This provides a more precise and flexible notion of the space of EFTs, a consistent way to encode the data needed to describe them, and it is essential for formulating the global conjecture. Moreover, EFT coverings also offer a natural perspective on \emph{counting} effective theories: at fixed cutoff $\Lambda$, one can count the number of EFT descriptions by the number of domains in a covering and make this notion canonical by minimizing over all coverings. More generally, one can refine this counting by assigning weights to domains, e.g.~by a suitable volume measure on each region as recently suggested in \cite{Baykara:2025nnc}. The Finite Complexity Conjecture then ensures that these counting prescriptions are mathematically well-defined and finite, and in particular that for EFTs valid at least up to a fixed cutoff $\Lambda$ the number of domains in an EFT covering and their tame complexities are uniformly bounded by functions of~$\Lambda$.

The paper is organized as follows. In section~\ref{sec:tameness} we review tameness via o-minimality and introduce its quantitative refinement, sharp o-minimality, which assigns a tame complexity $(\cF,\cD)$ to sets and functions. We then explain how this framework can be used to defined a complexity of a QFT, emphasizing a Lagrangian-based viewpoint and illustrating how complexity controls geometric and computational features. In section~\ref{sec:conjecture} we extend these ideas to EFTs: we discuss how apparently infinitely many Wilson coefficients can be reorganized into finite complexity descriptions, incorporate parameter/moduli dependence, and introduce EFT domains and finite EFT coverings to capture the need for multiple local descriptions over parameter space. In section~\ref{sec:complexity and quantum gravity}, we apply this formalism to study  the set of EFTs consistent with quantum gravity and we formulate local and global versions of the Finite Complexity Conjecture for these theories, relate them to swampland finiteness principles, and provide supporting evidence from string compactifications. We also explore consequences for the geometry of moduli/parameter spaces, including implications for volume growth and for quantitative notions of counting EFTs and vacua. We conclude in section~\ref{sec:conclusion} with an outlook and collect technical volume estimates in appendix~\ref{app_volumebounds}.

\section{Tameness, complexity, and first applications 
}\label{sec:tameness}

In this section we will introduce the formalism of tameness, made precise by the concept of o-minimality, which provides a solid logical framework for the notions of finiteness and complexity that we will later apply to effective field theories. This formalism, born from mathematical model theory, has experienced a rapid development in the recent years, including applications to physics. We will start by reviewing the original ideas of tameness, then we will discuss its quantitative refinements and most recent proposals and finish by establishing a first contact with quantum field theories.

\subsection{O-minimality}
\paragraph*{Defining o-minimal structures.}
We focus on stating the main ideas of o-minimality, and we refer to the foundational book \cite{VdDries} for a detailed exposition. The key idea behind o-minimality is to formalize what it means for a geometric object  to be tame. 
This is done by demanding that these objects are defined from a restricted selection of sets, known as an \textit{o-minimal structure}. Such a structure consists of a collection $\cS=(\cS_1,\cS_2,\ldots)$, where each $\cS_n$ is itself a collection of subsets of $\bbR^n$, obeying the following axioms:
\begin{enumerate}
    \item[(i)] $\cS$ is closed under finite unions and intersections;
    \item[(ii)] $\cS$ is closed under linear projections, complements, and products;
    \item[(iii)] $\cS$ contains all zero sets of polynomials;
    \item[(iv)] all sets in $\cS_1$ must have finitely many connected components. 
\end{enumerate}

The first point ensures that the structure is compatible with basic logical and geometric operations. It is required by the second point that all algebraic sets are included in the structure, so that tame geometry is at least an extension of algebraic geometry. Finally, the third point is the crucial o-minimality axiom which imposes a finiteness condition on the sets in $\cS_1$. Axiom (ii) ensures that this one-dimensional finiteness requirement persists in higher dimensions. The sets in an o-minimal structure $\cS$ are said to be definable in $\cS$, and we also refer to them as \textit{tame sets}. Additionally, a \textit{tame function} is a function whose graph is a tame set. 

\paragraph*{Triangulation and cell decomposition.}
The axioms given above together ensure that tame sets and functions have finitely many logical and geometric features. This is made precise by a number of powerful tameness theorems \cite{VdDries}. A striking example is the triangulation theorem, which states that any tame set is homeomorphic to a finite simplicial complex. This means that the topology of tame sets can always be obtained by gluing together finitely many points, lines, triangles, and higher-dimensional simplices, which already implies that all homology and cohomology groups of a tame set are finite-dimensional.

\begin{figure}[h!]
    \centering
    \includegraphics[width=\linewidth]{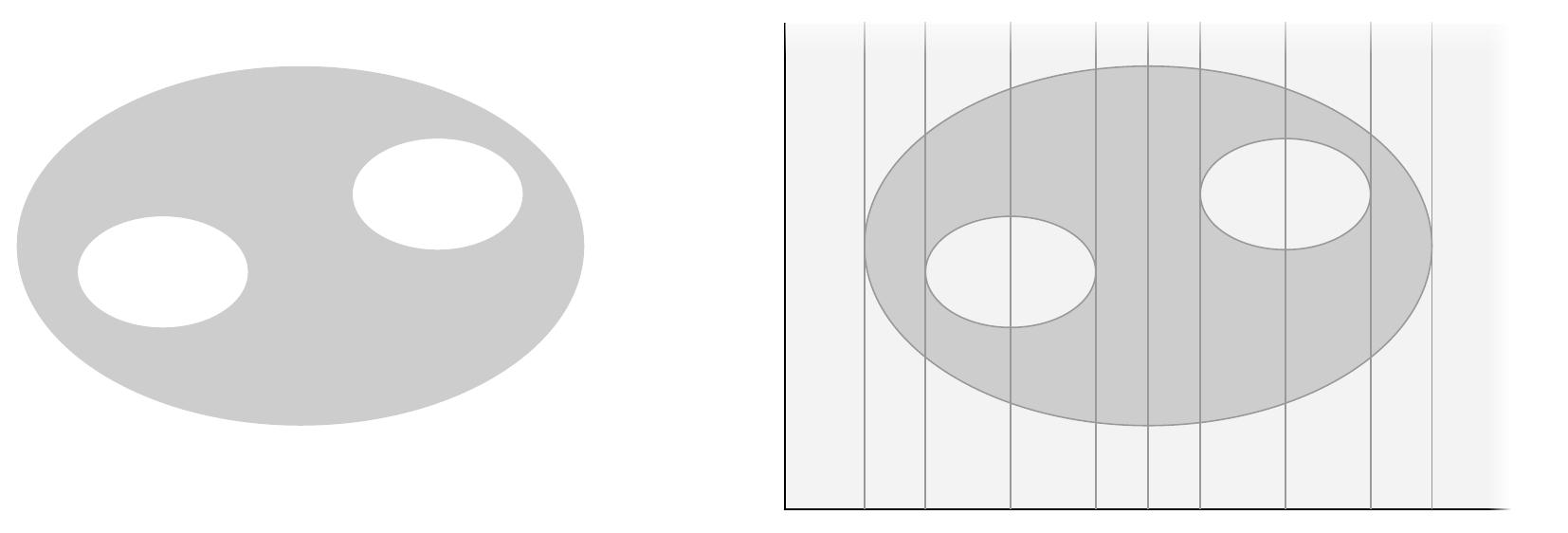}
    \caption{An example of a tame set $X\subseteq \bbR^2$ (left) and a cylindrical cell decomposition of $\bbR^2$ adapted to $X$ (right). The vertical boundaries of the cells ensure that the cell decomposition behaves appropriately under the linear projection to the horizontal axis.}
    \label{fig:CD}
\end{figure}

Another example is the cell decomposition theorem, which states that any tame set can be decomposed into elementary building blocks called cylindrical cells \cite{VdDries}. While the precise definition of cells is fairly technical, the idea behind them is simple: cells are essentially (hyper)cubes whose faces are parametrized by tame functions, constructed via the inductive property that linear projections of cells constitute cells themselves. An example is given in figure \ref{fig:CD}, which shows a tame set $X$ in $\bbR$ and a cell decomposition of $\bbR^2$, such that $X$ is a union of finitely many cells. The cell decomposition is chosen such that the boundaries of the two-dimensional cells are given by monotonic functions.

Theorems of this type make o-minimality a powerful setting for tame geometry, going far beyond algebraic geometry while still retaining some of its rigidity, yet excluding mathematically wild phenomena that appear in more general geometric settings.

\paragraph{Examples of o-minimal structures.}
The simplest o-minimal structure is called $\bbR_{\rm alg}$, which consists of semi-algebraic sets, which are subsets of Euclidean space cut out by polynomial equations and inequalities \cite{Tarski}. More general o-minimal structures are usually obtained by extending the class of polynomials to a function class $\cF$ consisting of sufficiently tame functions. One then considers the collection of sets cut out by equations and inequalities with functions from $\cF$, generating a structure denoted by $\bbR_\cF$. Examples of functions which generate o-minimal structures include the real exponential function \cite{WilkieExp} and analytic functions restricted to compact sets \cite{gabrielov1968projections}.

\paragraph{Tame geometry.}
The definition of o-minimality provides a tameness axiom for subsets of $\bbR^n$ for any $n$. It is possible to go beyond this, and generalize to a framework of tame geometry consisting of spaces and functions which are locally modeled after tame subsets of Euclidean space. In this spirit, a \textit{tame manifold} is a topological space which admits an atlas of \textit{finitely many} coordinate charts, such that the transition maps between the charts are tame functions. The requirement of finiteness of charts ensures that the geometry of tame manifolds retains the global finiteness features of tame subsets in Euclidean space. 

In this way it becomes possible to straightforwardly implement o-minimality in the standard geometric constructions on manifolds, such as vector bundles, differential forms, and Riemannian metrics, by demanding that they are tame when viewed through the tame coordinate charts on the manifold. This is especially important for applications to physics, where we frequently encounter spaces which are not naturally embedded in $\bbR^n$.

\subsection{Sharp o-minimality and complexity} 
\label{sharp-o-complexity}

Recently, a refinement of o-minimality was introduced, with the aim of quantifying the tameness of sets and functions. This is achieved by assigning a complexity to definable sets, based on the amount of logical information required to specify them. This refinement is known as sharp o-minimality. The inspiration for this framework is rooted in algebraic geometry, where the complexity of algebraic sets can be characterized by features such as dimension and degree. In particular, geometric and computational features of algebraic varieties depend polynomially on the degree, whereas the dependence on the dimension or number of variables is typically exponential or super-exponential. The key example of this fact is Bézout's bound, which states that the zero set of a degree $D$ polynomial with $N$ variables has at most $D^N$ connected components.

\paragraph{Defining sharply o-minimal structures.} Sharp o-minimality generalizes this idea by also characterizing the complexity of tame sets by two numbers, called \textit{format} $\cF$ and \textit{degree} $\cD$. The degree of a tame set can be regarded as the polynomial component of its complexity, whereas the format controls the remaining non-polynomial part. For a given o-minimal structure $\cS$, this is formalized by organizing the tame sets into a collection $\Omega=\{\Omega_{\cF,\cD}\}$ indexed by two integers $\cF,\cD\in\mathbb{N}$, which forms a filtration in the sense that 
\begin{equation}
    \Omega_{\cF,\cD}\subseteq \Omega_{\cF,\cD+1}, \quad \Omega_{\cF,\cD}\subseteq\Omega_{\cF+1,\cD}, \quad \text{and}\quad\bigcup_{\cF,\cD}\Omega_{\cF,\cD}=\cS\,.
\end{equation}
The collection $\Omega$ is called an \textit{FD-filtration}, and the subsets in a component $\Omega_{\cF,\cD}$ are thought of as having complexity $(\cF,\cD)$. In order to define a meaningful notion of complexity, we have to demand certain consistency requirements with the axioms of o-minimality:
\begin{enumerate}
    \item[(i)] \textbf{(Unions and intersections)} If $A_i \in \Omega_{\cF_i,\cD_i}$ for $i=1,\ldots,k$ and $A_i\subseteq \bbR^n$, then
     \begin{equation}
     \bigcup_{i=1}^ k A_i ,\,\,  \bigcap_{i=1}^ k A_i \in \Omega_{\cF,\cD} \,,
     \end{equation}
     where $\cF=\max_i\{ \cF_i\}$ and $\cD=\sum_i \cD_i$;
    \item[(ii)] \textbf{(Projections, complements, and products)} If $A \in \Omega_{\cF,\cD}$ and $A\subseteq \bbR^n$, then 
     \begin{equation}
     \pi(A), \,\, \bbR^n\setminus A  \in \Omega_{\cF,\cD} \,,
     \end{equation}
     for any linear projection $\pi:\bbR^n\to\bbR^{n-1}$, and
     \begin{equation}
     A\times \bbR, \,\,\bbR\times A \in\Omega_{\cF+1,\cD} \,.
     \end{equation}
    \item[(iii)] \textbf{(Algebraic sets and dimension)} for any $n$-variable polynomial $P$ of degree $m$, 
    \begin{equation}
        \{P=0\} \in \Omega_{n,m} \,.
    \end{equation}
    Moreover, if $A \in \Omega_{\cF,\cD}$ and $\bbR^n$, then $\cF \geq n$.
    \item[(iv)] \textbf{(Sharp o-minimality)} If $A\in\Omega_{\cF,\cD}$ and $A\subseteq \bbR$, then $A$ has at most $P_{\cF}(\cD)$ connected components, where $P_{\cF}(\cD)$ is a fixed polynomial in $\cD$ depending on $\cF$.
\end{enumerate}
The conditions (i) and (ii) control how the complexity of tame sets behaves under basic set-theoretic operations. In particular, under unions and intersections the formats are maximized and the degrees add, while linear projections and complements preserve format. Condition (iii) implements the idea that format and degree generalize the dimension and degree from algebraic geometry. Finally, axiom (iv) is the crucial axiom of sharp o-minimality which makes tameness quantitative. It assumes that the o-minimal structure $\cS$ with its FD-filtration $\Omega$ comes equipped with a collection of polynomials $\{P_\cF\}_{\cF\in\mathbb{N}}$ which control the number of connected components of tame sets in terms of their format and degree $(\cF,\cD)$. We henceforth refer to the pair $(\cF,\cD)$ of a tame set as its \textit{tame complexity}, or simply \textit{complexity} when the context is clear.

\paragraph{Example -- Pfaffian functions.} 
Extending this notion in a meaningful way beyond algebraic sets is a highly non-trivial task, and the best-established example is given by the class of Pfaffian functions \cite{Khovanskii}. These are functions $f_1,\ldots,f_r:U\to \bbR$ defined by a special system of differential equations, called a Pfaffian chain, of the form
\begin{equation}
     \frac{\pd f_i}{\pd x_j} = P_{ij} (x_1,\ldots,x_n, f_1,\dots,f_i) \,,
     \label{eq: pfaffian chain}
\end{equation}
where $i=1,\ldots,r$ and $j=1,\ldots,n$, and $P_{ij}$ is a polynomial depending on $i+n$ variables. In particular, note that the derivative of $f_i$ depends on the variables $x_1,\ldots,x_n$ and the previous functions $f_1,\ldots,f_i$ in the chain. Given such a chain, a \textit{Pfaffian function} is a function of the form
\begin{equation}
    f(x_1,\ldots,x_r) = P(x_1,\ldots,x_n,f_1,\ldots,f_r)\,,
    \label{eq: pfaffian function}
\end{equation}
where $P$ is a polynomial. The triangular structure of Pfaffian chains gives this class of functions remarkably controlled computational properties \cite{GabVor04} which have been used to show that they generate an o-minimal structure $\bbR_{\text{Pfaff}}$ \cite{WilkieRPfaff}. This means that any set defined through solutions to equations involving Pfaffian functions is tame. Crucially, the computational and geometric properties of these functions can be quantified in terms of the parameters of the Pfaffian chain. In particular, a Pfaffian function $f$ can be assigned a format and degree given by 
\begin{align}
        \cF &= r+n \,, \\
        \cD &= \deg P + \sum_{i,j}\deg P_{ij} \,.
\end{align}
It is shown that geometric features of sets defined by these functions, such as Betti numbers, exhibit a polynomial growth in $\cD$ \cite{Fewnomials,GabVor04,EffCell}. For example, the number of connected components of the set $\{f=0\}$ defined by a Pfaffian function with complexity $(\cF,\cD)$ has at most 
\begin{equation}
   b_0(\{f=0\}) \leq  2^{\cF^2} \cF^\cF \cD^\cF
\end{equation}
connected components.

\paragraph{Log-Noetherian functions.}

Another example of a tame structure that has a strong relevance for physical applications are Log-Noetherian (LN) functions. This class of functions constitutes a natural complex-valued generalization of the Pfaffian setting that has been the source of recent mathematical development in the context of o-minimality \cite{binyamini2024lognoetherianfunctions}. Let us briefly discuss how to construct them and highlight the main new features that distinguish them from their Pfaffian counterparts.   

To formally define a Log-Noetherian function, first one needs to introduce a Log-Noetherian cell as a complex version of the cells discussed briefly in the previous section. Inside this cell, the Log-Noetherian function will admit an explicit well-defined representation in terms of a set of differential equations. An $n$-dimensional Log-Noetherian cell is then constructed inductively as a fibration over an $n-1$-dimensional Log-Noetherian cell, where the fibers are either discs, punctured discs, annuli, or points, and the radii are tame functions. On such a cell $\cC$, a Log-Noetherian function $f: \mathcal{C} \to \mathbb{C}$, is then constructed by  finding a system of differential equations known as a Log-Noetherian chain following an analogous procedure to the Pfaffian setting described in \eqref{eq: pfaffian chain} and \eqref{eq: pfaffian function}, with the main differences being that the triangularity condition is removed and that logarithmic singularities are allowed.

Log-Noetherian functions generate the o-minimal structure $\mathbb{R}_{\rm LN}$, which is conjectured to be sharply o-minimal. This is a major point of interest for physical applications \cite{Carrascal:2025vsc}, since Log-Noetherian functions include period integrals, which play a pivotal role in very diverse areas, from Feynman integrals to  geometric constructions in quantum gravity compactifications. While the proper FD-filtration is not currently known, important steps towards this goal have been recently taken in the form of an effective notion of complexity for $\mathbb{R}_{\rm LN}$, given by a single number known as the effective format \cite{binyamini2024lognoetherianfunctions}. For a function $f: \mathcal{C} \to \mathbb{C}$ defined from a chain as the polynomial $f = P(f_1,\dots,f_N)$,  it can be computed as follows
\begin{equation} \label{eq:format-ln-func}
    \mathcal{F}_{\rm eff}(f) = \mathcal{F}_{\rm eff}(f_1,\dots,f_N) + \text{deg } P + \Vert P \Vert\,,
\end{equation}
where we have kept the same notation as in the Pfaffian case and defined the norm $\Vert \cdot \Vert$ of a polynomial as the sum of the absolute values of its coefficients. The format of the chain is itself computed by 
\begin{equation}\label{formatLNchain}
    \mathcal{F}_{\rm eff}(f_1,\dots,f_N)  = \cF_{\rm eff}(\cC) + N + \sum_{k,l} \text{deg } P_{kl} + \Vert P_{kl} \Vert + \sup_{\substack{i=1, \dots, N\\ \mathbf{z}\in \mathcal{C}}} |f_i (\mathbf{z})|\,.
\end{equation}
Notably, the format of Log-Noetherian functions is much more sensitive to the domain of definition, as it gets a contribution coming from the supremum of the functions inside the chain and it also depends on the format of the cell $\cC$ itself.\footnote{In the simplest cases, when the cell consists of direct products of $n$ discs and punctured discs of uniform radii $\{r_k\}_{k=1}^n$, its format can be taken to be $\cF_{\rm eff}(\cC)=n+\sum_{k=1}^{n} \lceil r_k \rceil$,
where  $\lceil \cdot \rceil$ is the ceiling function. We refer to \cite{binyamini2024lognoetherianfunctions,Carrascal:2025vsc} for more details on the source of the different contributions and the more rigorous evaluation of the format of the cells.}

The main takeaway from this brief overview is that many of the functions that appear in physical processes, such as period integrals, admit a well defined notion of complexity that requires to decompose their domain into cells where the functions posses a suitable description in terms of solutions to differential equations. The resulting complexity is heavily dependent on the properties of the domain and its cell decomposition. This effective complexity is conjectured to be compatible with a sharp o-minimal structure and the link between both descriptions can be intuitively established by taken the slice $\cF=\cD(=\cF_{\rm eff})$ in the sharp o-minimal filtration. In particular, we expect the polynomial dependence on the radius coming through the choice of cell to contribute to the degree of the representation of a given function.

\paragraph{Representations and duality.} Any measure of complexity has inherent ambiguities caused by the choice of elementary building block and how to assign complexity to these building blocks. In sharp o-minimality, this is taken as a built-in feature through the FD-filtration $\Omega$: instead of assigning a unique complexity to a tame set $X$, it assigns a `spectrum' of complexities given by all pairs $(\cF,\cD)$ for which $X\in\Omega_{\cF,\cD}$, which automatically includes all possible different representations of $X$. In particular, the framework naturally incorporates dualities, by inherently including alternative descriptions of an object which may have a lower complexity. An example of this idea is the theory of \textit{fewnomials} \cite{Fewnomials}. Consider a monomial $f(x)=x^N$. Viewed as a polynomial, $f$ has complexity $(1,N)$. However, there exists an alternative description as a Pfaffian function, through the Pfaffian chain 
\begin{align}
    \frac{\pd g}{\pd x} &= -g^2 \,,\\
    \frac{\pd f}{\pd x} &= m f g \,.
\end{align}
In this representation, $f$ has complexity $(2+1,2+2)=(3,4)$ independent of $N$. More generally, if $f$ is a polynomial of degree $N$ with $M$ monomial terms and $M\ll N$, i.e.~a so-called fewnomial, then it has a representation as a Pfaffian chain of length $r = M+1 $ through which it has complexity $(M+2,2M)$. In this `dual' description the complexity is independent of $N$, which is particularly interesting in the large $N$ limit. 

\paragraph{Emergence of simplicity.} Another essential example which underlines the importance of duality in the complexity framework of sharp o-minimality is the idea of \textit{emergence of simplicity} noted in \cite{Grimm:2024elq}. This is the phenomenon in which a sequence of tame sets or functions $X_N$ has a tame complexity which diverges as $N\to\infty$, while the limiting object $X = \lim_{N\to\infty}X_N$ has finite tame complexity due to the emergence of a simpler dual description. As an example, consider the function $f(x)=\tanh(x)$ and its $N$th order Taylor expansion
\begin{equation}
    f_N(x) = \sum_{n=1}^N \frac{2^{2n}(2^{2n}-1)B_{2n} }{(2n)!}x^{2n-1}\,,
\end{equation}
where $B_{2n}$ is the $2n$th Bernoulli number. The function $f_N(x)$ is a polynomial with complexity $(1,N)$, which diverges as $N\to \infty$. 
However, in the $N\to\infty$ limit, the limiting function $f(x)$ has an emergent simpler description as a Pfaffian chain with finite tame complexity in the FD-filtration. This means that the FD-filtration of a sharply o-minimal structure naturally takes dual descriptions into account. This is a promising mathematical feature, since the emergence of simplicity appears naturally in physical settings, such as large $N$ limits in AdS/CFT.

\subsection{Towards complexity in quantum field theories}

In this subsection we briefly discuss the proposal of \cite{Grimm:2024elq} for defining the complexity of a quantum field theory. The main idea is to consider the tameness and complexity of the functions which characterize the theory, such as the Lagrangian or the scattering amplitudes. This is only valid if these functions are definable in a sharply o-minimal structure; otherwise they may be regarded as having infinite complexity. In \cite{Grimm:2024elq}, two alternative proposals to characterize the complexity of a field theory were presented: either through the description of the theory (in the form of the Lagrangian), or through the observables of the theory (in the form of amplitudes and correlation functions).

Each of the two perspectives underscores different fundamental aspects of the theory. Focusing on the Lagrangian gives a clear characterization of the complexity of the theory, which naturally grows when more fields and interactions are included in the theory. However, this perspective is based on the classical data and thus less sensitive to the quantum effects of the theory. For instance, it does not apply for QFTs which have no Lagrangian description. On the other hand, the bottom-up perspective of characterizing the complexity of a theory using its observables is valid in general and purely quantum in nature, but does not easily provide a global notion of complexity for the theory itself. With applications to EFTs in mind, we will focus on the Lagrangian perspective in the following, and use sharp o-minimality to assign a format and degree to a quantum field theory. 

\paragraph{Example -- potential and vacuum structure.} To illustrate how to compute the complexity in the Lagrangian description, we consider a simple but foundational example: a theory with $N$ real scalar fields and a general polynomial interaction, described by the following Lagrangian 
\begin{equation}\label{eq:Lscalars}
    \cL(\phi_1,\ldots,\phi_N) = -\frac{1}{2}\sum_{k=1}^N \pd_\mu \phi_k \, \pd^ \mu \phi_k - V(\phi)\,; \quad V(\phi)=\sum_{I,\,|I|\leq D} \lambda_I \phi^I\, , 
\end{equation}
where $I=(I_1,\ldots,I_N)$ is a multi-index, and we denote
\begin{equation}
    |I|=I_1+\cdots+I_N, \quad \lambda_I = \lambda_{I_1\cdots I_N}, \quad \phi^I = \phi_1^{I_1} \cdots \phi_N^{I_N}\, .
\end{equation}
We view the Lagrangian as a tame function on the $(d+1)N$-dimensional space spanned by the field variables $\phi_1,\ldots,\phi_N$ and their spacetime derivatives. Since $\cL$ is polynomial, its complexity is simply given by $(\cF,\cD) = ((d+1)N,D)$, showing a straightforward dependence on the spacetime dimension, the number of fields, and the degree of the interactions. 

Let us highlight a simple application of sharp o-minimality. Suppose that  we wish to solve a computational problem in this theory, such as estimating the number of critical points of the potential $V$. The computational complexity of this problem should then depend on the complexity $(\cF,\cD)=(N,D)$ of the potential. Indeed, Bézout's theorem gives a sharp upper bound
\begin{equation}\label{eq:Bezout}
 \text{number of critical points} \leq (\cD-1)^\cF  \,,
\end{equation}
regardless of the precise form of the potential.

The power of sharp o-minimality is that it goes far beyond this simple polynomial example, and it has the potential to describe complicated geometric functions, including periods of Calabi-Yau manifolds \cite{binyamini2022sharply,binyamini2024lognoetherianfunctions}. In these more complicated settings, the bound given in equation \eqref{eq:Bezout} generalizes to a polynomial $P_\cF(\cD)$ in $\cD$ depending on $\cF$, and there exists a similar bound for any computational or geometric problem which can be formulated using first-order logic.\footnote{ Connections between computational complexity and tameness where studied in the context of quantum gravity in \cite{Lanza:2023vee,Lanza:2024mqp,Lanza:2025qfz}. In particular, in \cite{Lanza:2025qfz} it was  shown that the complexity of the process of verifying the weak gravity conjecture grows exponentially in the number of gauge fields. This  agrees with the prediction from sharp o-minimality, since the number of gauge fields contributes to the format, so that polynomial growth is not expected.} For example, if the potential were a Pfaffian function of complexity $(\cF,\cD)$, the number of vacua would be bounded by $2^{\cF^2}\cF^\cF \cD^\cF$ \cite{Grimm:2024elq}. In the polynomial case, it would be feasible to find the vacua by hand, but for more complicated functions the universal bounds become very powerful. 

\paragraph{QFTs and renormalizability.}

The discussion above shows how the complexity of a theory can be characterized by the Lagrangian and how this can be applied to a computational problem. However, it neglects the quantum nature of the theory, and in particular and does not consider its renormalization, which would require the inclusion of additional terms that would contribute to the complexity. In $d>2$, renormalizability restricts the Lagrangian to be polynomial of degree $\cD\leq 2d/(d-2)$. A non-renormalizable Lagrangian requires the addition of infinitely many counterterms, meaning that the theory in fact has infinite complexity.  Therefore, it is to be expected that non-renormalizable theories are generically incompatible with o-minimality at the quantum level. This issue can be circumvented by including an energy cutoff $\Lambda$ beyond which the theory is no longer valid, i.e.~to consider an EFT. The effective perspective comes with new technical challenges, which will be addressed in the next section.

\section{Complexity bounds on effective field theories}\label{sec:conjecture}
In the previous section we introduced the basic tools to characterize the complexity of mathematical objects and briefly illustrated how to apply them to the context of QFTs. The aim of this section is to implement this idea for general EFTs, which will require us to deal with various technical aspects such as the complexity of infinite sums, cutoffs and parameter dependence. Eventually, the goal is to develop a consistent complexity framework capable of describing EFTs in terms of sharp o-minimality, which in particular will allow us to systematically capture and characterize the finiteness properties of EFTs in the quantum gravity landscape, as we discuss in section \ref{sec:complexity and quantum gravity}. In order to achieve this, we have to address two essential conceptual points:  
\begin{enumerate}
    \item[(1)] The Lagrangian of an EFT in principle contains all local operators consistent with symmetries constructible from the fields and their derivatives. 
    \item[(2)] EFTs frequently come in families, depending on a set of parameters which also contribute to the overall complexity of the effective theory. 
\end{enumerate}
By considering these challenges we will develop a deeper understanding about how the information of an  EFT is stored, how this information gets repackaged as the cutoff evolves and how different descriptions are needed to fully characterize a family of EFTs across its parameter space.

\subsection{Complexity and EFT Lagrangians}\label{ss: complex eft lagrangian}
Consider a general $d$-dimensional EFT Lagrangian
\begin{equation}
\cL = \sum_{n} \Lambda^{d-\Delta_n}c_n \cO_n(\Phi,\pd \Phi,\ldots) \,,    
\end{equation}
where $\Phi$ collectively denotes the field content, $\Lambda$ is the cutoff scale, and $\cO_n$ is a local dimension-$\Delta_n$ operator with (dimensionless) Wilson coefficient $c_n$. It appears at first that such a Lagrangian generically has infinite complexity, since it requires infinitely many independent Wilson coefficients to be specified. However, sharp o-minimality provides a resolution: in special cases, there may be algebraic relations among the set $\{c_n\}$ such that there are only finitely many independent coefficients. In these cases, the Lagrangian, as a function of $\Phi$, could potentially be resummed into a function which is definable in a sharply o-minimal structure with finite tame complexity. 

Fully uncovering this structure requires a non-perturbative understanding of the theory under consideration. In the following we will consider a few non-perturbative examples where the exact effective Lagrangian is available, showing that it is in principle possible for these functions to have finite tame complexity. We focus on the part of the Lagrangian which only depends on the fields, i.e.~the potential, and discuss higher-derivative interactions at the end of the subsection.

\paragraph{Zero-dimensional QFT.} In zero dimensions, path integrals reduce to ordinary integrals, which enables exact calculations that can shed light on how effective actions may be described with finite complexity. We will illustrate this by considering a simple 0d QFT with two scalars, described by the Euclidean Lagrangian
\begin{equation}
    \cL(\phi,\chi) = \frac{1}{2} m^2 \phi^2 + \frac{1}{2} M^2 \chi^2 + \frac{\lambda}{4!} (\phi^2+ \chi^2)^2\,.
\end{equation}
Integrating out the $\chi$ field yields an effective Lagrangian for $\phi$, which is given by
\begin{equation}
    \cL_{\rm eff}(\phi) = \sum_{n\geq 0}  c_{2n} \phi^ {2n} = \frac{1}{2} m^2 \phi^2 + \frac{\lambda}{4!}\phi^4 -\log\left[\int \dd\chi \, e^ {-\frac{1}{2} (M^2+ \lambda\phi^2/6 )\chi^2 + \frac{\lambda}{4!} \chi^2} \right] \,,
\end{equation}
with an infinite set of Wilson coefficients $c_{2n}$. These coefficients can be obtained perturbatively by expanding the interaction term inside the integral over $\chi$, and collecting the contributions to each effective interaction in $\phi$. Performing this procedure, we find
\begin{equation}
    c_0 =- \log(Z_0)\,,  \quad  c_2 = \frac{1}{2}m^2 + a_2\,, \quad  c_4 = \frac{\lambda}{4!} + a_4 \quad\text{and} \quad c_{2n}= a_{2n} \quad \text{for  }n\geq3\,, 
\end{equation}
where for $n\geq 1$ the sequence $a_{2n}$ can be expressed as
\begin{equation}
     a_{2n} =  \sum_{\ell=1}^n  \tfrac{1 }{\ell} \big(- \tfrac{2}{Z_0}\big)^ \ell \!\! \sum_{   \substack{k_1+\cdots+k_\ell =n\\ k_i\geq 1}} \prod_{i=1}^ \ell \Big(  \sum_{k=0}^ \infty     \tfrac{ (-1)^ {k+k_i} }{ k!k_i! }   \left( \tfrac{ \lambda }{6}   \right)^{k+k_i}M^{-4k-2k_i-1} \,\Gamma\big(2k+k_i+\tfrac{1}{2}\big)  \Big) \,,
\end{equation}
and $Z_0$ is the partition function of $\chi$ evaluated at $\phi=0$, 
\begin{equation}
    Z_0 = \int \dd\chi \, e^ {-\frac{1}{2} M^2\chi^2 + \frac{\lambda}{4!} \chi^4} \,.
\end{equation}
Note that the expression for $a_{2n}$ is written in terms of asymptotic series due to the divergent nature of the perturbative expansion. 

The expressions for the Wilson coefficients are rather complicated, and it is not immediately apparent how the resulting effective Lagrangian $\cL_{\rm eff}(\phi)$ could have a finite tame complexity. However, the 0d path integral can be performed exactly, yielding an analytic expression for the effective Lagrangian given by 
\begin{equation}
    \cL_{\rm eff}(\phi) = \frac{1}{2} m^2 \phi^2 + \frac{\lambda}{4!}\phi^ 4 -\log \left[  \sqrt{\frac{3}{\lambda}  \mu^2(\phi)    }  \exp(\frac{3\mu^4(\phi) }{4\lambda}) K_{1/4}\left( \frac{3\mu^4(\phi)}{4\lambda} \right)      \right] \,,
\end{equation}
where $\mu^2(\phi) = M^2 +\tfrac{1}{6}\lambda\phi^2$ is the effective squared mass of the $\chi$ field, and $K_{1/4}$ is a modified Bessel function of the second kind. It was shown in \cite{Grimm:2023xqy,Grimm:2024elq} that this function can be described in terms of a Pfaffian chain of differential equations, meaning that it is definable in the o-minimal structure $\bbR_{\rm Pfaff}$, and in particular that it has finite tame complexity. This simple example illustrates the following point: even if the effective Lagrangian has infinitely many independent Wilson coefficients, each of which takes a complicated form, it can still potentially be resummed into an object of finite complexity, provided that the UV theory is sufficiently simple and under computational control. The finiteness of complexity relies on a differential equation obeyed by the effective action, which reorganizes the information content of the infinite set of Wilson coefficients into a finite amount of information. The tameness of non-perturbative partition functions in more general 0d QFTs was studied in \cite{Grimm:2024hdx}.

\paragraph{Exact renormalization group in higher-dimensional EFTs.} Ultimately, we are interested in the complexity of EFT Lagrangians in $d\geq 1$. In this case, exact calculations are rare, and more powerful techniques are required. In the following we will discuss an approach using exact renormalization group techniques. The resulting RG flows are notoriously difficult to solve, but here sharp o-minimality offers the advantage that the complexity of an object can already be extracted from an implicit description such as a differential equation, and an exact closed-form expression is not needed.

Consider a $d$-dimensional EFT describing a single scalar field $\phi$ with a cutoff energy scale $\Lambda_0$, with a Euclidean action
 \begin{equation}
     S_{\Lambda_0}[\phi] = \int \dd^ d x  \left(\frac{1}{2}(\pd \phi)^2 + V_{\Lambda_0}(\phi) \right) \,,
 \end{equation}
 where $V_{\Lambda_0}(\phi)$ is the potential of the theory at the scale $\Lambda_0$. Suppose that we lower the energy scale to $\Lambda<\Lambda_0$ by integrating out all modes with energy between $\Lambda$ and $\Lambda_0$. Formally, the resulting effective potential $V_{\Lambda}(\phi)$ is obtained by performing the path integral over modes within this energy range. Denoting the lower-energy modes by $\phi$ and the modes with energy between $\Lambda$ and $\Lambda_0$ by $\chi$, the Wilsonian effective action becomes
  \begin{equation}\label{eq:wilsonpathint}
     S_{\Lambda}[\phi] = -\log \left[ \int D \chi \, e^ {-S_{\Lambda_0}[\phi+\chi]}   \right]\,,
 \end{equation}
 from which the effective potential $V_{\Lambda}(\phi)$ can be extracted.

To analyze the complexity of the effective potential, we continue in the local potential approximation (LPA), in which higher-derivative interactions generated by the path integral are projected out. This path integral can almost never be performed analytically, but upon lowering the scale infinitesimally the one-loop contribution becomes exact, and the effective potential obeys the non-perturbative Wegner-Houghton equation \cite{Wegner:1972ih}
\begin{equation}\label{eq:WegnerHoughton}
    \Lambda\frac{\pd}{ \pd \Lambda} V_{\Lambda}(\phi) = - \Lambda^d A_d \log\left(\Lambda^ 2 + \frac{\pd^2}{\pd\phi^2}V_{\Lambda}(\phi) \right) \,,
\end{equation}
where $A_d =  [(4\pi)^{d/2}\Gamma(d/2) ]^ {-1}$ is an angular integration factor (see e.g.~\cite{Aoki:1996fn,Bonanno:1999ik} for a discussion).
The fact that an infinite tower of quantum corrections to the effective potential can be captured by a single differential equation suggests the compelling idea that there may be a finite-complexity description of the effective theory which is definable in an o-minimal structure. More precisely, we expect that a tame initial potential $V_{\Lambda_0}(\phi)$ flows under this equation to a tame effective potential $V_{\Lambda}(\phi)$ for $\Lambda<\Lambda_0$. The structure of equation \eqref{eq:WegnerHoughton} suggests that the corresponding o-minimal structure could be the one generated by Log-Noetherian functions \cite{binyamini2024lognoetherianfunctions} or the Pfaffian extension \cite{Speisegger99} of this structure. This idea of conservation of tameness is further supported by the observation that RG flow behaves qualitatively similar to heat diffusion. Precise details on partial differential equations of this type and o-minimality are presently not known, and it is likely that a non-trivial mathematical extension is required to describe the tameness of exact RG flows. We leave a detailed analysis of these ideas for future work, but before we move on, it is interesting to note that non-perturbative renormalization group flow techniques \cite{Wegner:1972ih, Wetterich:1992yh} have also been developed in the context of quantum gravity effective theories \cite{Reuter:1996cp,Lauscher:2001ya,Reuter:2007rv,Saueressig:2023irs}, which provides further motivation to the proposal of section  \ref{sec:complexity and quantum gravity}.

\paragraph*{Supersymmetric theories -- Seiberg-Witten example.} In EFTs with supersymmetry, the structure of the effective Lagrangian is much more constrained, and  non-perturbative statements can often be inferred with the help of non-renormalization theorems and geometric constructions. 
This happens for instance for BPS states in supersymmetric theories, whose masses robustly survive quantum corrections. In some cases, higher perturbative corrections vanish completely due to the holomorphy or symmetry constraints and one can attain exact expressions of the Wilson coefficients of the effective theory, characterizing them completely with a finite amount of information.

We will illustrate this with a well-known example, namely Seiberg-Witten theory \cite{Seiberg:1994rs} (we refer to \cite{Seiberg:1994aj,klemm:1995wp, Lerche:1996xu,Alvarez-Gaume:1996ohl} for reviews). This theory is a geometric description of  $\text{SU}(2)$ pure gauge theory with $\mathcal{N}=2$ supersymmetry in four dimensions. In the absence of matter hypermultiplets, the effective theory can be decomposed into two $\mathcal{N}=1$ chiral multiplets $W_\alpha$ and $A$, and we denote by $a$ the scalar component of $A$. The presence of $\mathcal{N}=2$ supersymmetry allows us to express the effective Lagrangian in terms of a single holomorphic function $F(A)$, the prepotential, as follows:
\begin{equation}
\begin{aligned}
        \mathcal{L} = & \frac{1}{4\pi} \mathrm{Im} \Bigg[ 
\int \dd^2\theta \,\dd^2\bar{\theta} \,\frac{\partial F(A)}{\partial A} \bar{A} + \int \dd^2\theta \, \Big( \frac{1}{2} \sum_{\alpha} \frac{\partial^2 F(A)}{\partial A^2 } W^{\alpha} W_{\alpha} \Big) 
\Bigg]\,.
\end{aligned}
\label{eq: SW lagrangian}
\end{equation}
The quantum corrections to the classical prepotential $F=\frac{1}{2}\tau_0 a^2$ were found to be perturbatively exact at one-loop order, with non-perturbative corrections from a series of instantons \cite{Seiberg:1988ur}. The full expression thus takes the form
    \begin{equation}
    F(a)=\frac{1}{2}\tau_0 a^2+\frac{i}{\pi}a^2\log\left(\frac{a}{\Lambda}\right) +\frac{1}{2\pi i}a^2\sum_{n=1}^\infty c_n \left(\frac{\Lambda}{a}\right)^{4n}\,,
    \label{eq: SW prepotential}
    \end{equation}
where the $c_{n}$ parametrize the instanton corrections and $\Lambda$ is a dynamically generated scale. The main realization of \cite{Seiberg:1994rs, Seiberg:1994aj} was to identify the moduli space of $\text{SU}(2)$ Yang-Mills with that of an elliptic curve, which allows for an exact calculation of the coefficients $c_n$. This description thereby yields full control over the non-perturbative effective action. It was shown in \cite{Matone:1995rx} that they can be encoded in terms of the function $G(a)=\pi i \big(F(a)-\frac{1}{2}a\partial_a F(a)\big)$, which satisfies the differential equation 
\begin{equation}
    (1-G^2)\frac{\dd^2G}{\dd a^2}+\frac{a}{4}\left(\frac{\dd G}{\dd a}\right)^3=0\,.
\end{equation}
These two differential equations allow one to write the prepotential $F$ as a Log-Noetherian function as the ones described in section  \ref{sec:tameness}, thereby making the effective action \eqref{eq: SW lagrangian} definable in an o-minimal structure, and giving it a description with finite complexity.\footnote{Note that this goes beyond the analysis of \cite{Grimm:2023xqy}, where it was shown that the Seiberg-Witten gauge coupling $\tau = \pd_a^2 F(a)$ has finite complexity when restricted to a real half-line in the moduli space.}\footnote{To fully describe the Lagrangian in terms of Log-Noetherian functions over the whole moduli space, one needs to cover the moduli space by special regions called cells. We will study this point more in depth in section \ref{sec:parameters}.} From this differential equation, one can recover the coefficients $c_n$ by means of a recursion relation \cite{Matone:1995rx}, which further supports the viewpoint that infinitely many a priori independent coefficients can be captured with finite information. 

This alternative description of the differential equations that govern Seiberg-Witten theory helps to intuitively illustrate how the infinite series expansion of the prepotential can have finite complexity. In \cite{Carrascal:2025vsc} a detailed global characterization of the complexity of the coupling function (i.e. the period map $\tau=\partial^2 \mathcal{F}/\partial a^2$) was provided using an effective o-minimal framework \cite{binyamini2024lognoetherianfunctions}. We will discuss more about this example and its potential generalization to quantum gravity effective theories in the following sections.

\paragraph*{Geometry of higher-derivative interactions.}
So far, we have focused on the potential of the EFT. However, a general EFT Lagrangian contains an infinite series of higher-derivative interactions. In order to consider derivative interactions in the framework of o-minimality, which in principle only describes functions and sets of finitely many real variables, we have to introduce an auxiliary real variable for each spacetime derivative. For example, in a theory with a scalar $\phi$ we introduce auxiliary variables $X_\mu = \pd_\mu \phi$, $X_{\mu\nu}=\pd_{\mu}\pd_{\nu}\phi$, etc. so that a higher-derivative operator in the Lagrangian becomes a monomial in the auxiliary variables, e.g.
\begin{equation}
    \phi^{p} \big (\pd_\mu\phi  \,\pd^ {\mu }\phi\big)^ {q} \big(\pd_{\nu}\pd_\rho\phi \, \pd^\nu \pd^\rho \phi  \big)^{r} \longrightarrow \phi^{p} (X_\mu X^ \mu)^ {q} (X_{\nu\rho} X^ {\nu\rho} )^{r} \,.
\end{equation}
Recently, this approach was implemented for EFT Lagrangians in \cite{Craig:2023wni,Craig:2023hhp}, where the Lagrangian was formulated as a function on the jet bundle over field space, parametrized by the fields and the auxiliary derivative variables. In particular, it was argued in \cite{Craig:2023hhp} that the jet bundle EFT Lagrangian has a great amount of geometric structure which persists to arbitrarily high order in the derivative expansion.

\paragraph{Complexity of the higher-derivative expansion.} In this formulation, it is a well-defined mathematical question to consider the tameness and complexity of the EFT Lagrangian as a function of the fields and the auxiliary variables including higher-derivative interactions. The Wegner-Houghton equation can be systematically modified beyond the local potential approximation to describe the non-perturbative RG evolution of higher-derivative interactions \cite{Bonanno:1999ik}, so that the discussion earlier in this section extends to higher derivatives. 

When extending this discussion to the full EFT Lagrangian including all orders in the derivative expansion, this approach runs into a fundamental problem for tameness. The expansion would require the inclusion of infinitely many auxiliary variables, i.e.~one would have to formulate the Lagrangian on the infinite jet bundle, while o-minimality is only defined in finite-dimensional settings. One way of addressing this issue would be to truncate the derivative expansion, introducing only finitely many auxiliary variables, and consider the complexity of the truncated EFT Lagrangian. Note that one can then still consider general functions of these auxiliary variables, which means that effectively some of the infinite towers of derivative interactions remain. For example, if one includes the first and second derivatives of the spacetime metric as auxiliary variables, one can still consider the complexity of general functions $f(R,R_{\mu\nu},R_{\mu\nu\rho\sigma})$ of the curvature invariants in a gravitational EFT. An alternative approach would be to formulate the Lagrangian in momentum space, in which the replacement $\pd_{\mu}\to p_\mu$ completely circumvents the introduction of auxiliary variables. We could then consider the complexity of the function $\cL(\widetilde\Phi,p_\mu)$, where $\widetilde\Phi$ collectively denotes the Fourier transforms of the field content. This approach would in principle allow one to quantify the complexity of the complete EFT expansion. Since we focus on the conventional position-space Lagrangians, we will not consider this approach here. In the remainder of this work, \textit{we focus mainly on the two-derivative sector of the EFT Lagrangian}.

\subsection{Complexity, parameter spaces, and EFT domains}\label{sec:parameters}
In the previous subsection, we have focused on the complexity of the EFT Lagrangian as a function of the fields. We now extend this discussion to also include the complexity associated to the external parameters. Consider an EFT with a single scalar field $\phi$, whose two-derivative Lagrangian takes the general form 
\begin{equation}
    \cL(\phi,\pd\phi) = -\frac{1}{2}\pd_\mu \phi\, \pd^ \mu \phi + \sum_{n=0}^\infty c_n \phi^n \,.
\end{equation}
The complete parameter space of this family of EFTs is the infinite-dimensional space $\cM$ spanned by the Wilson coefficients $\{c_n\}$. As we have learned in the previous subsection, a generic point in this space determines an EFT with infinite complexity. However, on special subsets of $\cM$, the Lagrangian may be recast into a tame function of finite tame complexity. With this idea in mind, we can formally cover $\cM$ by subsets on which the Lagrangian has a \textit{fixed} complexity. We will call these sets \textit{EFT domains}, and formalize this concept at the end of this section.

\paragraph{Example -- complexity and symmetries in 0d.}
To illustrate this notion in a simple example, consider the Lagrangian
\begin{equation}\label{eq:SON-Lagrangian}
    \cL(\phi_1,\ldots,\phi_N) = \frac{1}{2}\sum_{j=1}^N m_j^2 \phi_j ^2  +  \frac{\lambda}{4!} \Big(\sum_{j=1}^N  \phi_j^2\Big ) ^2 \,,
\end{equation}
describing a 0d theory of $N$ scalars $\phi_1,\ldots,\phi_N$ with a quartic interaction. The parameter space $\cM$ of this family of theories is the $(N+1)$-dimensional space given by the values of $m_1^2,\ldots,m_N^2$ and $\lambda$. At a generic point in $\cM$, the tame complexity of the Lagrangian is $(\cF,\cD)= (N,4)$. The subset given by $\lambda=0$ then defines an $N$-dimensional EFT domain on which the complexity reduces to $(N,2)$. Consider now the loci on which $n$ masses coincide, e.g.~$m_1^2=\cdots = m_n^2=m^2$. Within this set, the theory acquires an $\text{SO}(n)$ symmetry, and by rewriting the fields $\phi_1,\ldots,\phi_n$ in spherical coordinates, they can be fully described in terms of a radial field $\rho$. Here the Lagrangian becomes\footnote{Note that at the quantum level, there is an additional term $(N-1)\log \rho$ in the effective Lagrangian due to the Jacobian of the path integral measure $\dd \phi_1\cdots\dd\phi_N =\rho^{N-1} \dd \rho \, \dd \Omega_{N-1}$. This term can be included as a Pfaffian function, and increases the complexity slightly but independently of $N$ and $n$.} 
\begin{equation}
    \cL(\rho,\phi_{n+1},\ldots,\phi_N) = \frac{1}{2} m^2\rho^2 + \frac{1}{2}\sum_{j=n+1}^N m_j^2 \phi_j ^2   +  \frac{\lambda}{4!} \Big(\rho^2+\sum_{j=n+1}^N  \phi_j^2\Big ) ^2\,,
\end{equation}
so that this subset of the parameter space defines an EFT domain on which the Lagrangian has reduced complexity $(N-n+1,4)$. The parameter space $\cM$ can in this way be fully decomposed into EFT domains with varying complexity. In this example, it is clear that the EFT covering of $\cM$ is associated to symmetries of the Lagrangian.

\begin{figure}[h!]
    \centering
    \includegraphics[width=0.95\linewidth]{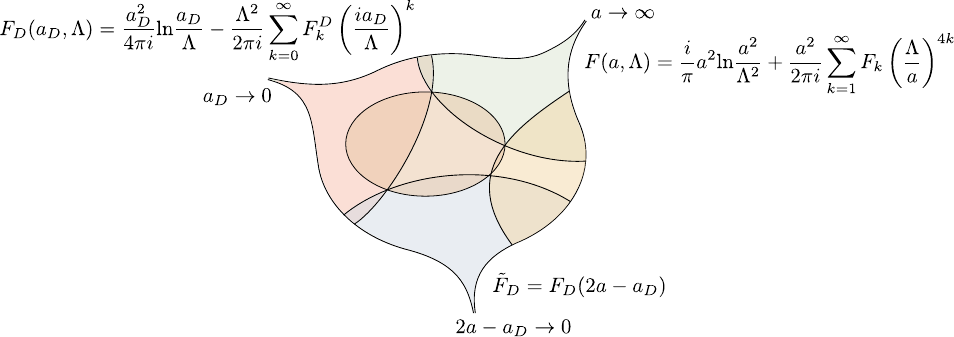}
    \caption{EFT covering of $\mathcal{N}=2$ $SU(2)$ Seiberg-Witten moduli space with finitely many EFT domains (discs and punctured discs). Around each of the three infinite distant limits there exists a different EFT description of finite complexity in terms of different combinations of variables $a$ and $a_D$, capturing the electric-magnetic duality discussed in \cite{Seiberg:1994rs}. In addition, three other discs centered around regular points are included to cover the full moduli space.}
    \label{fig: SW patches}
\end{figure}

\paragraph{EFT domains and local Lagrangians.} In the discussion above, the parameter space was taken to be the space of couplings or a subset thereof, and there was a globally valid Lagrangian over the entire parameter space. In many cases, for instance for the EFTs arising from quantum gravity, the parameter space of a family of EFTs instead consists of a set of vacua formed by vacuum expectation values (vevs) of fields, and the values of the Wilson coefficients depend on these vevs. In these cases, an EFT covering of the parameter space is further constrained by the fact that there may not be a globally valid Lagrangian. For example, consider the $\text{SU}(2)$ Seiberg-Witten theory, for which the Coulomb branch moduli space coincides with the moduli space of an elliptic curve and it is given  by a Riemann sphere with three marked points. In this setting it is known that there is no globally valid EFT Lagrangian \cite{Seiberg:1994aj}. Instead, the moduli space must be covered by at least three regions, each located around a different singularity and endowed with their own locally valid EFT description, as depicted in figure \ref{fig: SW patches}. In this case, the underlying physical mechanism is the convergence of the instanton expansions around the various singularities. The complexity of the Lagrangians of Seiberg-Witten theory, including the EFT covering of the moduli space, is studied in great detail in \cite{Carrascal:2025vsc}.

\paragraph{Geometry of moduli space.} Finally, let us note that the parameter space $\cM$ of an EFT can be a complicated manifold which is not naturally embedded as a subset of Euclidean space. In order to implement the framework of sharp o-minimality, we must therefore view $\cM$ as a tame manifold, and assign a complexity $(\cF,\cD)$ through the tame coordinate charts. Hence, in order to define the complexity of a locally valid Lagrangian in parameter space, the notion of EFT domain must be compatible with the tame coordinates on $\cM$.

\paragraph{Definition of EFT coverings.} The discussion of this section suggests that it is useful to introduce a notion of EFT domains and EFT coverings. These concepts resonate with Hardt’s decomposition theorem for definable families \cite{VdDries} (see section~\ref{sec:tameness} for comments on the cell decomposition).  More precisely, given a set of EFTs parametrized by a space $\cM$, we define an \textit{EFT domain} as a subset $U\subseteq \cM$ on which the EFTs are described by a single Lagrangian $\cL_U$ with a fixed complexity $(\cF_U,\cD_U)$. In other words, viewing the Lagrangian of the theory as defining the fibers, all fibers can be identified over $U$. 
An \textit{EFT covering} $\mathcal{U}$ is a collection of EFT domains $U_i$ which cover $\cM = \bigcup_i U_i$. We require that they have the following properties:
\begin{itemize}
    \item Each EFT domain has a well-defined dimension. Different domains might have different dimensions. In contrast to the cells of a tame set, we allow the $U_i$ to overlap. 
    \item Every EFT domain $U$ is assumed to have a tame coordinate map $U\to V\subseteq \bbR^n$, and the complexity of the Lagrangian is evaluated on the set $V$.
    \item On regions in the parameter space where there is no Lagrangian with finite complexity, we formally set $\cF$ and $\cD$ to infinity. 
\end{itemize}
Intuitively, an EFT covering is simply a parametrization of Lagrangian descriptions which together cover the parameter space of a set of EFTs.

\section{Complexity bounds and quantum gravity}
\label{sec:complexity and quantum gravity}

In this section, we build on the ideas of section \ref{sec:conjecture}
and combine them with constraints arising from quantum gravity to study the complexity of effective field theories in the quantum gravity landscape. This will lead us in section~\ref{complexity_conjecture} to propose two  Finite Complexity Conjectures on EFTs in the quantum gravity landscape. In section~\ref{complexity_comp} we provide further motivation for these conjectures by discussing how finite complexity EFTs can arise in string theory and, more generally, from compactification of a higher-dimensional theory. Finally, in section~\ref{sec:complexity_volume_counting} we discuss the connection of the Finite Complexity Conjectures with other swampland conjectures related to volumes in moduli spaces and counting of theories. 

\subsection{Swampland conjectures on complexity bounds} \label{complexity_conjecture}

We now turn to the formulation of the two conjectures. First we focus on an individual EFT, and propose a local version of the conjecture. We then take a more global perspective, and propose a stronger version of the conjecture which states that the complexity of EFTs is uniformly bounded across the landscape. We motivate the conjectures from general quantum gravity principles, and provide illustrative examples in 
section~\ref{complexity_comp}.

We have seen that completely generic EFT Lagrangians require an infinite amount of information to be described, encoded in the infinite set of Wilson coefficients. However, in top-down cases where the UV origin of the effective theory is sufficiently well-understood, either by non-perturbative techniques, supersymmetry, or other geometric constraints, it appears that this information can be reorganized into an object of finite complexity. This claim should hold, in particular, when considering effective theories that are compatible with quantum gravity, which is expected to impose various finiteness constraints that have been collected within the swampland program.

 \paragraph{Finiteness of light states and the swampland.} Let us make a clearer connection between the finiteness of complexity and finiteness constraints proposed in the swampland program. From the inception of the program, it was argued that quantum gravity bounds the ranks of gauge groups and massless spectra of EFTs \cite{Vafa:2005ui,Kumar:2009ae,Kumar:2010ru, Morrison:2011mb,Park:2011wv,Grimm:2012yq}; an idea that is supported by the claimed finiteness of compact Calabi-Yau threefolds \cite{reid1987moduli,yau2008survey}, which has seen much progress when restricting to elliptic fibrations \cite{Gross1993AFT, MR4939522,MR4801611,Birkar:2025gvs}. In particular, bounding the field content in the setting of six-dimensional supergravity has recently been a very active topic ~\cite{Lee:2019skh,Kim:2019vuc,Katz:2020ewz,Tarazi:2021duw,Kim:2024eoa,Birkar:2025rcg} (see also \cite{Martucci:2022krl} for a discussion on bounds in four dimensions). The finiteness of massless and light states is further supported by the general expectation that quantum gravity imposes a fundamental cutoff scale $\Lambda_{\rm QG}$ beyond which the EFT description is no longer valid. Depending on the mass hierarchies of the fundamental states of the underlying UV theory, this scale can have a different origin as has been investigated in many recent works \cite{Dvali:2007hz,Dvali:2007wp,Dvali:2008ec,Grimm:2018ohb,Castellano:2021mmx,vandeHeisteeg:2022btw,Cribiori:2022nke,Castellano:2022bvr,vandeHeisteeg:2023dlw,Blumenhagen:2023yws,Burgess:2023pnk,Bedroya:2024uva,Calderon-Infante:2025ldq,ValeixoBento:2025bmv}. 
A universal upper bound on $\Lambda_{\rm QG}$ is expected to be given by the species scale $\Lambda_{\rm sp}$, which is parametrically lowered when the number of light species increases. Consequently, EFTs coupled to gravity cease to be valid when too many light species are included. In particular, for a given EFT cutoff $\Lambda$, we must have  $\Lambda<\Lambda_{\rm QG}$, which bounds the number of fields. These two ideas together imply that the contribution to the complexity of the Lagrangian coming from the number of fields is bounded for EFTs in the landscape. The suggested spectrum finiteness is non-trivial for individual EFTs but becomes much stronger when claimed to hold uniformly for the set of all EFTs.   

\paragraph{Finiteness of complexity -- individual EFTs.} Even if one assumes the finiteness of states in the EFT, the Lagrangian could still depend on infinitely many independent Wilson coefficients. Supporting our expectation to have a finite complexity representation, reference \cite{Heckman:2019bzm} argues that the Wilson coefficients of an EFT consistent with quantum gravity are determined by finitely many parameters. This implies that the infinitely many terms of the EFT Lagrangian are actually related to each other, hinting towards the existence of a finite complexity representation of the EFT expansion as discussed in sections~\ref{sharp-o-complexity} and~\ref{sec:conjecture}.\footnote{A typical way to encode relations among the Wilson coefficients is to describe the series with a differential equation with finitely many controllable coefficients as exemplified for Pfaffian structures in section~\ref{sharp-o-complexity}.} 
Taken together, these observations lead us to propose a local version of the Finite Complexity Conjecture:

\vspace{0.2em}
\noindent\fbox{
\parbox{\dimexpr
\linewidth-2\fboxsep-2\fboxrule}
{\vspace*{.1cm}
\noindent \textbf{Local Finite Complexity Conjecture.} 
Every EFT consistent with quantum gravity has a description in which the two-derivative Lagrangian is definable in a sharply o-minimal structure and has finite tame complexity $(\cF_{\rm EFT},\cD_{\rm EFT})$.\\[-.3cm]
}}
\smallskip

\noindent

Let us emphasize that this conjecture is formulated at the level of an individual EFT in the quantum gravity landscape. It formalizes the idea that each of these EFTs is non-generic and admits a description of finite tame complexity $(\cF_{\rm EFT},\cD_{\rm EFT})$, with format and degree depending on the considered EFT.
It can be viewed as a quantitative refinement of the claim that the Lagrangian of an  EFT consistent with quantum gravity is a tame function, upgrading the corresponding part of the Tameness Conjecture \cite{Grimm:2021vpn} by replacing o-minimality with sharp o-minimality. The conjecture is also strictly stronger than the finiteness claim for Wilson coefficients proposed in \cite{Heckman:2019bzm}, since functions depending on only finitely many parameters may still be non-tame while appearing to carry finite information. The distinction arises because sharp o-minimality imposes finiteness constraints that are stable under all logical operations, rather than merely at the level of parametrization.

It is also important to highlight that the conjecture is formulated only for effective theories admitting a Lagrangian description, and that the evaluation of the complexity is obtained by restricting the Lagrangian to the two-derivative level. Rather than being motivated by physical arguments, these conditions arise from the limitations of the application of the formalism of sharp o-minimality to quantum field theories detailed in sections \ref{sec:tameness} and \ref{sec:conjecture}. However, the picture is by no means complete or unique, and alternative approaches could allow for a more general characterization of complexity that enable an extension of the regime of validity of the conjecture. In particular, we expect that including a finite number of higher-order derivative fields by the procedure explained in the previous section would increase the complexity bounds while maintaining the core statement of the conjecture. Accounting for the complete higher-derivative expansion is much more challenging and requires developing new procedures for the evaluation of the complexity, as discussed at the end of subsection \ref{ss: complex eft lagrangian}. Finally, extending the validity of the conjecture to non-Lagrangian theories, though beyond the scope of this work, may be possible by implementing the complexity framework in a different way. The consistency with quantum gravity still imposes finiteness constraints on these non-Lagrangian theories, and we expect that generality of sharp o-minimality makes it possible to formulate appropriate finite complexity bounds for these theories, for instance by analyzing their observables \cite{Douglas:2022ynw,Douglas:2023fcg ,Grimm:2023xqy}, geometric aspects of their moduli spaces, or dualities relating them to Lagrangian theories. We leave this extension for future research.

\paragraph{Finiteness of complexity -- landscape of EFTs} 

While the local finite complexity conjecture proposes that each effective field theory in the landscape admits a finite, EFT-dependent complexity $(\cF_{\rm EFT},\cD_{\rm EFT})$, it does not exclude the possibility that there exist families of consistent quantum gravitational EFTs whose complexity becomes unbounded. However, it is widely believed that the landscape of EFTs arising from string theory—and more generally from any consistent theory of quantum gravity—is finite once suitable cutoffs are imposed \cite{DouglasStringsTalk,Vafa:2005ui,Acharya:2006zw,Hamada:2021yxy}.

For example, it has been conjectured in \cite{Hamada:2021yxy} that there are only finitely many distinct EFTs consistent with quantum gravity that are valid up to a chosen cutoff scale $\Lambda$ if one identifies two EFTs in the same connected moduli space. Such finiteness statements are supported by the aforementioned broad body of results \cite{Kumar:2009ae,Kumar:2010ru, Morrison:2011mb,Park:2011wv,Grimm:2012yq,reid1987moduli,yau2008survey,Gross1993AFT,MR4939522,MR4801611,Birkar:2025gvs,Lee:2019skh,Kim:2019vuc,Katz:2020ewz,Tarazi:2021duw,Kim:2024eoa,Birkar:2025rcg,Martucci:2022krl} as well as by the recent general proofs of the finiteness of flux vacua in Type IIB string theory and F-theory \cite{Bakker:2023xkt,Grimm:2021vpn}. A first conceptual refinement of these finiteness results was formulated in the Tameness Conjectures \cite{Grimm:2021vpn,Douglas:2023fcg}, which asserts that the field spaces and parameter spaces of EFTs valid below a cutoff $\Lambda$ are definable in an o-minimal structure. While this guarantees finiteness in a qualitative sense, it does not assign quantitative bounds and therefore does not yet allow for explicit, computable measures of complexity.

Extending the Tameness Conjectures of \cite{Grimm:2021vpn,Douglas:2023fcg} and the Local Complexity Conjecture stated above, we arrive at the hypothesis that there is a uniform bound on the complexity of EFTs in the landscape. We formalize this in our main conjecture as follows:

\vspace{0.2em}
\noindent\fbox{
\parbox{\dimexpr
\linewidth-2\fboxsep-2\fboxrule}
{
\vspace*{.1cm}
\noindent \textbf{Finite Complexity Conjecture.} For fixed spacetime dimension $d$, consider a finite energy cutoff $\Lambda>0$ in $d$-dimensional Planck units.
\begin{itemize} 
\item[(i)] The set of $d$-dimensional EFTs consistent with quantum gravity with energy cutoff at least $\Lambda$ is definable in a sharply o-minimal structure and has finite tame complexity $(\cF_\Lambda,\cD_{\Lambda})$.

\item[(ii)] There exists a pair of positive integers $(\mathfrak{F}_\Lambda,\mathfrak{D}_\Lambda)$, such that any $d$-dimensional EFT consistent with quantum gravity with energy cutoff at least $\Lambda$ has a description for which the two-derivative Lagrangian is definable in a sharply o-minimal structure and has a tame complexity $(\cF_{\rm EFT},\cD_{\rm EFT})$ satisfying $\cF_{\rm EFT}\leq \mathfrak{F}_\Lambda$ and $\cD_{\rm EFT}\leq\mathfrak{D}_\Lambda$.
\end{itemize}
\vspace*{-.1cm}
}}

\vspace*{.1cm}

\noindent
Informally speaking, the conjecture proposes that there is an information limit on effective theories of quantum gravity, implemented by a uniform complexity bound across the landscape. It integrates various perspectives on finiteness ideas that have been suggested in the swampland program and opens the possibility to give quantitative bounds.  
Let us now discuss the statement of the conjecture in more detail.

\paragraph{Part (i) of the conjecture.} Let us begin by discussing the first part of the conjecture. It concerns a set $\cM_{\rm QG;\Lambda}$ which is assumed to parameterize the space of EFT Lagrangians that are valid at least up to some cutoff scale $\Lambda$. In general, the space $\cM_{{\rm QG};\Lambda}$ is expected to be a very complicated object, consisting of a formal union of parameter spaces, moduli spaces, and field spaces of the EFTs. Its precise definition will require to address several important points, e.g.~when two EFTs are considered to be equivalent. Furthermore, the structure and boundaries of $\cM_{{\rm QG};\Lambda}$ will be enforced by inherent consistency of the theories and in particular by the requirement that the considered EFTs are compatible with quantum gravity. The conjecture that the full space $\cM_{\rm QG; \Lambda}$ is definable in a sharply o-minimal structure therefore constitutes a great challenge to verify in generality. However, the complexity constraints translate also to all smaller subspaces obtained, for example, by appropriate restrictions.\footnote{Note that this is going to the heart of sharp o-minimality, where finiteness claims and the existence of the complexity measures are compatible with operations such as linear projections or intersections.} This fact allows for testing the conjecture in many settings or finding concrete counter-examples. We will highlight some basic examples in favor of the conjecture in section~\ref{complexity_comp}. 

We also stress that even though the conjecture claims that $\cM_{{\rm QG};\Lambda}$ has finite complexity, it leaves open how precisely one assigns the complexity to this set. In principle, a format and degree can only be assigned to sharply o-minimal subsets of Euclidean space. If $\cM_{{\rm QG};\Lambda}$ has more structure, e.g.~parts that are Riemannian manifolds, it is tempting to introduce a notion of sharply o-minimal manifolds. A natural definition of such a manifold is obtained by 
tracing through the definition of a tame manifold, and quantifying the complexity of each step in the construction. In this way, the complexity would correspond to the sum of the complexities of the tame coordinate regions and their coordinate transition functions. 
An alternative perspective is to embed $\cM_{{\rm QG};\Lambda}$ into Euclidean space, so that it admits directly a well-defined format and degree $(\cF_{\Lambda}^{\rm emb},\cD_{\Lambda}^{\rm emb})$. At first, the additional requirement of an embedding may appear unnatural. However, if one combines the embedding with the idea that field spaces have an intrinsic notion of distance, and one takes this metric into account by demanding that the embedding is isometric, one recovers various known quantum gravity constraints \cite{Grimm:2025lip}. We will come back to these interpretations in 
section~\ref{sec:complexity_volume_counting}, where we will also comment on how complexity is related to volume growth and counting. 

\paragraph{Moduli space of quantum gravity theories.} Given our notion of $\cM_{{\rm QG};\Lambda}$, we now introduce the set of all EFTs consistent with quantum gravity denoted by $\cM_{\rm QG}$ as being given by the $\Lambda\to 0$ limit of $\cM_{{\rm QG};\Lambda}$. This limiting procedure picks up all EFTs consistent with quantum gravity, valid up to an arbitrarily small $\Lambda$. We can try to assign a complexity $(\cF_{\rm QG},\cD_{\rm QG})$ to $\cM_{\rm QG}$ by a limiting procedure, if we are able to find representations for all $\cM_{{\rm QG};\Lambda}$ that have a uniform bound $\cF_{\Lambda} \leq \cF_{\rm QG}$ and $\cD_{\Lambda}  \leq \cD_{\rm QG}$.\footnote{This is the familiar notion of so-called limit sets in tame geometry \cite{van2005limit}.} It is then tempting to extend the part (i) of the conjecture to the whole space $\cM_{\rm QG}$ and claim that $(\cF_{\rm QG},\cD_{\rm QG})$ are finite. It turns out, however, that one then would have to refine the notion of $\cM_{\rm QG}$, since otherwise there are immediate counter-examples arising when the parameter set is infinite and discrete, such as for Type IIB string theory backgrounds like AdS$_5\times S^5$ supported by $N$ units of flux (see section~\ref{complexity_comp}). We will discuss a possible way to deal with these cases in section~\ref{sec:complexity_volume_counting} and refrain from extending the conjecture at this point. 

It is also instructive to compare our definition of $\cM_{\rm QG}$ with the notion of moduli spaces used in \cite{Ooguri:2006in} and many subsequent follow-ups. In string theory settings one often introduces a moduli 
space $\cM$ without setting a cutoff scale, and therefore does not work within a single EFT. In our picture, we view these moduli spaces $\cM$ as components of $\cM_{\rm QG}$, which might be labeled by additional (potentially discrete) parameters. This allows us 
to extend the conjectures about viable moduli spaces $\cM$ to $\cM_{\rm QG}$. In particular, extending \cite{Ooguri:2006in} one obtains the statement that $\cM_{\rm QG}$ is parametrized by inequivalent expectation values of fields and there are no other free parameters.\footnote{Note that this is a familiar fact within string theory but it is far from apparent there is no a priori reason this should be the case in every theory of quantum gravity.} In section~\ref{sec:complexity_volume_counting} we
will further comment on how some of the swampland conjectures apply to $\cM_{\rm QG}$ and interplay with the tameness of $\cM_{{\rm QG};\Lambda}$. 

\paragraph{Part (ii) of the conjecture.} The tame geometry of $\cM_{{\rm QG};\Lambda}$ and the Local Finite Complexity Conjecture, imply that the parameter space $\cM_{{\rm QG};\Lambda}$ has a \textit{finite} EFT covering, i.e.~a covering by finitely many domains in which an EFT Lagrangian can be defined. This formalizes and generalizes the conjecture on the finiteness of EFTs valid up to a cutoff $\Lambda$:

\vspace{0.2em}
\noindent\fbox{
\parbox{\dimexpr
\linewidth-2\fboxsep-2\fboxrule}
{
\vspace*{.1cm}
\noindent \textbf{Finiteness of EFTs.} The parameter space $\cM_{{\rm QG};\Lambda}$ of EFTs with cutoff at least $\Lambda$ has a finite EFT covering. 
\vspace*{.1cm}
}}

\vspace*{.1cm}

\noindent
In particular, for a given $\Lambda$, there should in principle be a mininum number $\cN_\Lambda$ of EFT domains required to cover $\cM_{{\rm QG};\Lambda}$, which would depend on the complexity $(\cF_\Lambda,\cD_\Lambda)$ as well as on the complexity of the corresponding EFTs. The part (ii) of the conjecture then implies the all of these EFTs have a common upper bound on their complexity $(\mathfrak{F}_\Lambda,\mathfrak{D}_\Lambda)$.

Assuming that the explicit values for $(\mathfrak{F}_{\Lambda},\mathfrak{D}_\Lambda)$ are known, any EFT with cutoff $\Lambda$ that violates these bounds by not having any description with $\cF_{\rm EFT}<\mathfrak{F}_\Lambda$ and $\cD_{\rm EFT}<\mathfrak{D}_\Lambda$ belongs to the swampland.

\subsection{Finite complexity from string theory and compactification} \label{complexity_comp}

In this section we discuss a number of examples supporting the Finite Complexity Conjecture. 
In order to do that, we start from string theory as a candidate theory of quantum gravity and explore the landscape of its low-energy EFTs. In addition to supporting the statement of the conjecture, some of the examples will 
clarify why the assumptions of the conjecture are necessary. 

\paragraph{Scales in quantum gravity.} 
Before turning to explicit examples, let us discuss in more detail the role of the cutoff scale. String theory generically contains infinite towers of excitations -- string modes, and in lower-dimensional compactifications, additional towers such as Kaluza–Klein states. Since an EFT of finite complexity cannot include infinitely many individual states, one must introduce a cutoff $\Lambda$ that restricts the effective description to finitely many degrees of freedom. Validity of the EFT requires $\Lambda\le \Lambda_{\rm QG}$, where $\Lambda_{\rm QG}$ denotes the intrinsic quantum gravity scale at which the EFT breaks down. Determining the appropriate estimate for $\Lambda_{\rm QG}$ is therefore important for the Finite Complexity Conjecture. The cutoff is below the $d$-dimensional Planck mass $M_{\rm pl}$ and minimally has to account for the impact of light species, leading to the species scale $\Lambda_{\rm sp}<M_{\rm pl}$ \cite{Dvali:2007hz,Dvali:2007wp,Dvali:2008ec}. However, further obstructions can arise \cite{Burgess:2023pnk,Bedroya:2024uva}. In the presence of an interacting tower of states, such as a Kaluza–Klein tower, a natural choice is the mass of the lightest state in the tower, defining the tower scale $\Lambda_{\rm t}$. This is well-motivated near infinite-distance limits of moduli space, where towers are expected to become light \cite{Ooguri:2006in}, but $\Lambda_{\rm t}$ might fail to control quantum-gravitational effects in the bulk. To address this, the black hole scale $\Lambda_{\rm BH}$ was introduced in \cite{Bedroya:2024uva}, signaling EFT breakdown through black hole instabilities. It satisfies $\Lambda_{\rm BH}\le \Lambda_{\rm sp}$ and approaches $\Lambda_{\rm t}$ in asymptotic regions. In the examples below we focus on the asymptotic regime and adopt $\Lambda_{\rm t}$ as a natural cutoff (avoiding partial truncations of the tower); using $\Lambda_{\rm BH}$ would mainly modify the numerical complexity estimates while leaving the qualitative conclusions unchanged.

\paragraph{Ten-dimensional supergravity theories.} Famously, the five consistent supersymmetric string theories reduce to five supergravity theories with $\cN=2$ (Type IIA/Type IIB) and $\cN=1$ (heterotic/Type I) supersymmetry. These theories have no free parameters other than the ten-dimensional Planck mass. Nevertheless they admit non-trivial moduli spaces $\cM_{\Lambda}$. We will now briefly discuss how these theories accept a description of finite complexity using Type IIA and Type IIB as examples.

\begin{figure}[h!]
    \centering
    \includegraphics[width=0.7\linewidth]{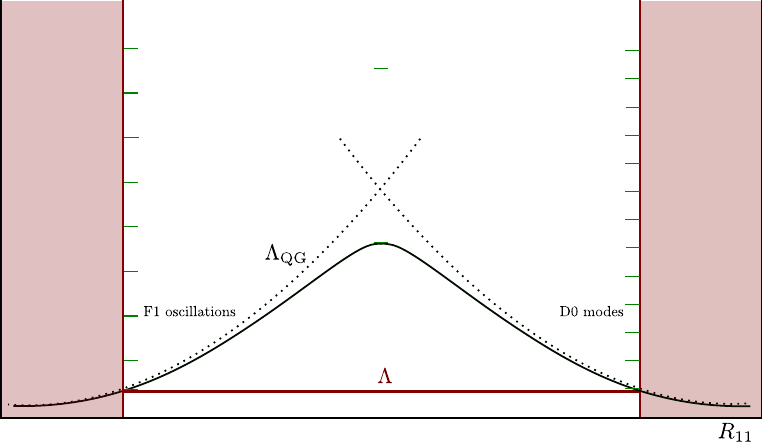}
    \caption{Depiction of the moduli space $\cM_{\Lambda}$ of Type IIA supergravity in 10 dimensions. The boundaries of the region are located at the points in which $\Lambda_{\rm QG}$ (given by in this case by the lightest state of the tower) falls below $\Lambda$. The shaded areas correspond to inaccessible regions of the moduli space of the compactification that do not admit an EFT with the chosen cutoff. }
    \label{fig: diagram IIA in 10d}
\end{figure}

The starting point to understand Type IIA is 11-dimensional supergravity, a theory with no scalars and no free parameters. Upon compactification on a circle, we obtain Type IIA supergravity in ten dimensions. The resulting theory has a single scalar, the dilaton, $\phi$, that is related to the radius of the compactifying circle through $R_{11}=e^{\frac{2}{3} \phi}$ in 11-dimensional Planck units. Consequently, the moduli spaces $\cM_{\Lambda}$ are finite-length intervals parameterizing the vacuum expectation values $g_s$ of the dilaton field $e^{\phi}$. The action of the massless fields including both fermionic and bosonic components can be found for instance in \cite{Bergshoeff:2001pv} up to four fermionic terms. It is possible to verify that both the field content and the functions involved have finite tame complexity. Even if we account for higher fermionic interactions, these will be always polynomial and thus their complexity will still be bounded \cite{Grimm:2024elq}. The massless spectrum of the theory is complemented by towers of massive states,  generated by string excitations (obtained from M2 branes wrapping the circle in the 11-dimensional picture) and the non-perturbative D0 states (which are the Kaluza-Klein modes of the compactification). The masses of these excited states are given by

\begin{equation}
     M^2=\frac{k}{R_{11}^2}+\frac{2}{\alpha'}(N_L+N_R+E_0)\,,
    \label{eq: mass excited states}
\end{equation}
where $k$ and $(N_L,N_R)$ count the Kaluza-Klein momentum and the string excitations respectively. From \eqref{eq: mass excited states}, it is clear that the cutoff $\Lambda_{\rm t}$, set by the mass of the lightest field in the towers, changes for every point of the moduli space spanned by $e^\phi$.\footnote{For constant Planck mass this means that $\alpha'$ changes as we move in moduli space since the $d$-dimensional string coupling, the $d$-dimensional Planck mass and the string mass are related by $1/g_{s,d}\sim \alpha'^2 M_{{\rm pl},d}^4$.} The complexity of the EFT becomes finite when considering a fixed cutoff $\Lambda$. If $\Lambda<\Lambda_{\rm t}$,
we are only left with the massless content and the associated Lagrangian can be written in terms of sharply o-minimal functions with bounded complexity, in agreement with the Finite Complexity Conjecture. Above $\Lambda_{\rm t}$, no EFT description exists.  Therefore, for a given choice of probe cutoff $\Lambda$, there is a patch $\mathcal{M}_{\Lambda}$ in the radial moduli space bounded by the values at which the Kaluza-Klein modes (in the large radius limit) or the winding modes (in the small radius limit) become lighter than $\Lambda_{\rm QG}$, as seen in figure \ref{fig: diagram IIA in 10d}.  It is inside this patch where there exists a well-defined Lagrangian description with finite complexity and cutoff valid at least up to $\Lambda$.

\begin{figure}[h!]
    \centering
    \includegraphics[width=0.5\linewidth]{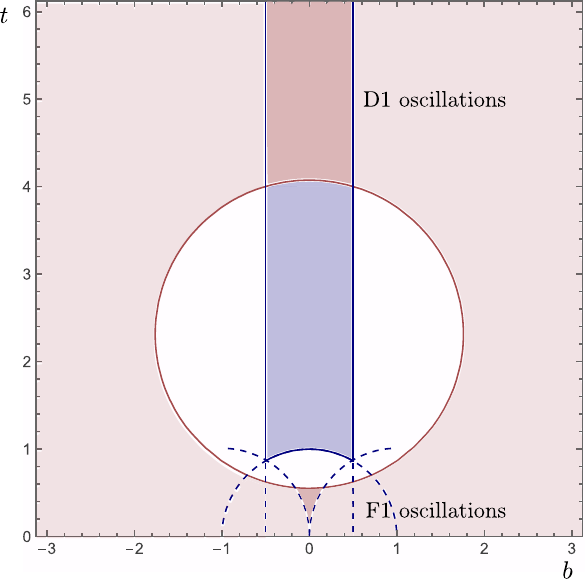}
    \caption{Depiction of the moduli space $\cM_{\Lambda}$ of Type IIB supergravity in 10 dimensions as a subset of the hyperbolic plane. The solid blue lines depict the boundaries of the fundamental domain of $\mathbb{H}/\text{SL}(2,\mathbb{Z})$. The region of the fundamental domain where the EFT supergravity description is valid is highlighted with shaded blue while the dark shaded red corresponds to the region of the fundamental domain where the masses of the tower of D1 oscillations fall below the cutoff $\Lambda$. Finally, the region outside the fundamental domain where the masses of the S-dual string tower fall below the cutoff has also been highlighted.  }
    \label{fig: diagram IIB in 10d}
\end{figure}

Type IIB supergravity in ten dimensions displays many of the same general features but the realization is considerably different, since in addition to the dilaton there is another free scalar field: the RR 0-form $c_0$. Together they generate the 2-dimensional axio-dilaton moduli space $\cM_{\rm IIB} = \mathbb{H}/\text{SL}(2,\mathbb{Z})$, that is, the upper half-plane with the hyperbolic metric quotiented by the $\text{SL}(2,\mathbb{Z})$ duality group of the theory. Writing $\mathbb{H}$ as a group quotient this space can be expressed as 
\beq \label{IIBquotient}
   \cM_{\rm IIB} = \text{SL}(2,\mathbb{Z}) \backslash \text{SL}(2,\bbR)/\text{SO}(2)\ .
\eeq
As it was the case in the Type IIA discussion, the complexity of the Lagrangian of the massless content is finite and towers of states become exponentially light in the infinite distance limits of moduli space. When fixing a cutoff, we can determine a region of finite geodesic distance where our EFT description is valid. After accounting for the quotient with the duality there is a single infinite distance limit and the allowed region at a fixed cutoff takes the shape shaded in blue in figure \ref{fig: diagram IIB in 10d}. It is important to highlight that quotienting by the duality is required in order to obtain a description of finite complexity. More details of the complexity of $\cM_{\rm IIB}$ within an o-minimal structure were discussed in \cite{Grimm:2025lip,Carrascal:2025vsc}. Below we will return to the discussion of complexities of quotient spaces such as \eqref{IIBquotient}.

\paragraph{M-theory on a torus.}

\begin{figure}[h!]
    \centering
    \includegraphics[width=0.8
    \linewidth]{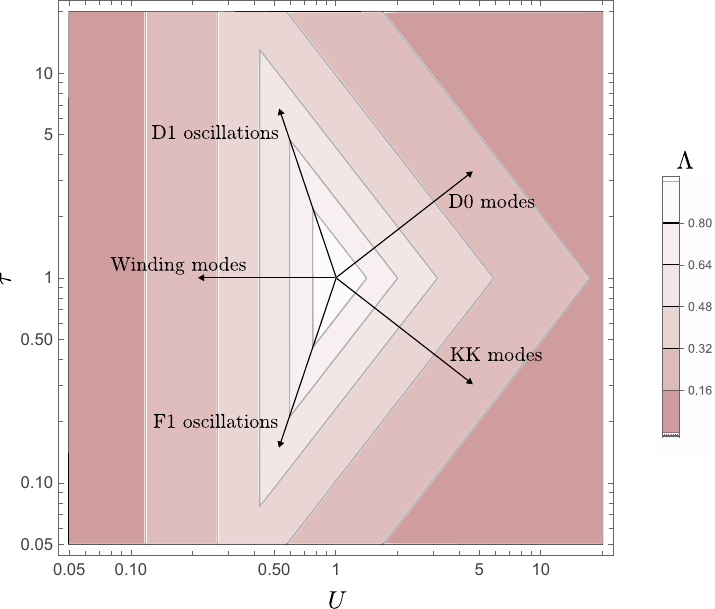}
    \caption{Shape of the moduli space $\cM_{\Lambda}$ of the 9-dimensional theory obtained from compactifying M-theory on a torus for different values of $\Lambda$ (in 9-dimensional Planck units). The moduli space has been parametrized using the volume of the torus $U=R_{10}R_{11}$ and its complex structure $\tau=R_{11}/R_{10}$. The towers whose light states set the boundary of validity of the effective theory along each direction are highlighted.}
    \label{fig: M torus moduli space}
\end{figure}

We can also consider theories in lower dimensions, starting with compactifications of Type IIA and Type IIB on a circle. Both of them can be discussed together as different limits of M-theory compactified on a torus. In this case, the moduli space becomes three dimensional, since the 9-dimensional massless spectrum now includes three scalars: the radius of the circle $R_{10}$, the lower-dimensional dilaton $\phi$ and its axionic partner (either the RR 0-form of Type IIB or the component along the compact dimension of the RR 1-form of Type IIA). The resulting moduli space is\cite{Schwarz:1995dk, Aspinwall:1995fw}
\begin{equation}
    \mathcal{M}_{T^2}=\mathbb{R}_{+}\times \text{SL}(2,\mathbb{Z}) \backslash \text{SL}(2,\bbR)/\text{SO}(2)\ .
\end{equation}
The complexity of the Lagrangian of the massless sector increases as the theory is compactified to lower dimensions, but it can still be bounded with an appropriate description. As in the previous examples, such description is only valid in a finite region of the moduli space, determined by the behavior of the massive towers of states along the different directions and a choice of cutoff. To see this more explicitly, note that now there are two non-compact directions to consider, namely the two radial directions of the toroidal compactification.\footnote{Since the axion does not play a prominent role in the asymptotic behavior of the towers, from this point onward we will assume it is fixed to a constant value, different choices being related through modular transformations \cite{Etheredge:2022opl,Calderon-Infante:2023ler,vandeHeisteeg:2023dlw}.} Along these two directions, in addition to the three towers of the 10-dimensional theories (F1, D1 and D0 modes) one must also add the Kaluza-Klein and winding towers that become light in the limits of large and small radius $R_{10}$ respectively. Together, for a fixed cutoff, they bound a compact region $\mathcal{M}_\Lambda$ of the two dimensional moduli space, as displayed in figure \ref{fig: M torus moduli space}. Outside of this region, the complexity of the starting 9-dimensional EFT diverges and the description is no longer valid.

These simple examples illustrate how the connected components of moduli spaces arising in quantum gravity can be covered by a finite number of patches -- each associated with an EFT description of finite complexity --  and how this covering is rooted on the network of dualities that permeates string theory (we will discuss more about this point at the end of section \ref{sec:complexity_volume_counting}). In more general cases we expect to require many more patches and have several distinct connected components, but the core underlying features will remain the same.

\paragraph{Complexity of (half-)maximal supergravity.} 
The EFT examples discussed so far are instances of supergravity theories with a large amount of supersymmetry. 
They can be viewed as part of the EFT landscape with maximal or half-maximal supersymmetry, which are particularly well controlled settings. In general, the Lagrangian of these theories is invariant under a semi-simple Lie group $G$ which is uniquely fixed by the spacetime dimension, number of supercharges, and number of vector multiplets \cite{CecottiBook}. Crucially, when such theories arise within quantum gravity, e.g.~via a string compactification on a simple manifold such as a torus, the theories must be quotiented by a duality group $\Gamma$ \cite{CecottiBook}. 
The scalar field spaces of these theories are \textit{arithmetic quotients}, which are spaces of the form
\begin{equation} \label{arithmeticq}
    \cM = \Gamma \backslash  G/ K \,,
\end{equation}
where $K\subseteq G$ is a connected compact subgroup, and $\Gamma\subseteq G$ is an arithmetic subgroup.\footnote{In the case of supergravity theories with more than 8 supercharges, $K$ is always a maximal compact subgroup.} Obviously, the example $\cM_{\rm IIB}$ discussed in \eqref{IIBquotient} is a special case of \eqref{arithmeticq}. It is a non-trivial result of \cite{BKT} that these spaces are tame manifolds definable in $\bbR_{\rm alg}$. Since the structure $\bbR_{\rm alg}$ is sharply o-minimal, this means that this field space has an explicitly computable finite complexity $(\cF,\cD)$ depending on $G,K,$ and $\Gamma$. In principle, this complexity may be calculated by counting the formats and degrees of the algebraic functions required to define the tame coordinate charts on $\cM$, following the construction of \cite{BKT} using Siegel sets \cite{BorelArGr}. 

Additionally, it was argued in \cite{Grimm:2021vpn,MickThesis} that, as a consequence of the invariance under the algebraic group $G$, the local Lagrangians of these supergravity theories are also definable in $\bbR_{\rm alg}$. Extending this to include the global duality group $\Gamma$ will likely not invalidate the finite complexity assertion for the Lagrangian, even though, as for $\cM$, we expect the arguments to be more involved. This then confirms the Finite Complexity Conjecture in this setting. One may go beyond this analysis and consider gauging a subgroup $G_0\subseteq G$, which turns the theory to a gauged supergravity theory and introduces a scalar potential on $\cM$. Given the algebraic nature of this procedure, it is a natural expectation that the local gauged supergravity Lagrangian, including the scalar potential, is still definable in $\bbR_{\rm alg}$. It is an interesting problem to check if the Finite Complexity Conjecture for such gauged supergravity theories with more than 8 supercharges can be established for all or only a subset of gaugings.

\paragraph{Compactification of Type IIB String Theory.} A next natural generalization is to look at more involved compactifications of string theory and test the Finite Complexity Conjecture for the arising EFTs. Such EFTs become quickly very complicated and, in the absence of supersymmetry or other controlling symmetry groups, quantum corrections modify the complexity computation significantly. However, in Type IIB string theory and F-theory compactifications on Calabi-Yau manifolds, we expect that at least certain sectors of the EFT admit computable finite complexity descriptions. In particular, many couplings in the subsector controlled by the complex structure moduli $\cM_{\rm cs}$
of a Calabi–Yau manifold are determined by the period integrals of the relevant Calabi–Yau forms over an integral basis of homology cycles. These integrals are encoded by the period map
\beq
   \Phi: \ \ \cM_{\rm cs} \ \rightarrow\ \Gamma \backslash  G/ K \ .
\eeq
One is therefore led to control the complexity of the moduli space $\cM_{\rm cs}$, the double quotient 
$\Gamma \backslash  G/ K $, which is in general no longer an arithmetic quotient, and the typically transcendental map $\Phi$. Despite these substantial challenges, it was conjectured in~\cite{beyondo-min,binyamini2024lognoetherianfunctions} that the period integrals are definable in a sharply o-minimal structure, allowing one to assign to them a well-defined complexity $(\cF,\cD)$. This conjecture constitutes a quantitative refinement of the tameness of period maps proven in  \cite{BKT}. It is expected that the o-minimal structure $\bbR_{\rm LN,exp}$ plays a central role in this analysis \cite{binyamini2024lognoetherianfunctions}. 
Establishing explicit bounds on the complexity of such effective field theories is likely to be highly non-trivial, and we expect that meaningful results can initially be obtained only for relatively simple examples. Building on the recent advances of \cite{binyamini2024lognoetherianfunctions,Carrascal:2025vsc}, it appears feasible to derive complexity bounds in an effective o-minimal sense for Calabi–Yau moduli spaces with a small number of moduli. While this framework should also accommodate additional sectors and known quantum corrections, a complete determination of the complexity of general string-theoretic EFTs currently seems out of reach.

\begin{figure}
    \centering
\includegraphics[width=0.65\linewidth]{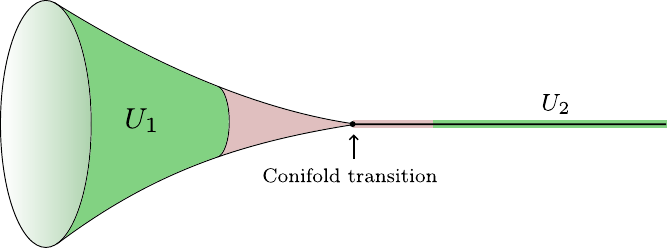}
    \caption{Depiction of the moduli space near a conifold transition between two topologically distinct compact Calabi-Yau manifolds. For a fixed cutoff, there are EFT domains $U_1$ and $U_2$ that provide 4-dimensional EFT descriptions of their respective compactifications near the conifold point without reaching it.  }
    \label{fig: conifold transition}
\end{figure}

Nevertheless, the complexity perspective provides a powerful lens through which to investigate quantitative properties of the string theory landscape. As noted above, the Finite Complexity Conjecture is naturally aligned with the expected finiteness of compact Calabi-Yau manifolds, reflecting the idea that putative discrete infinities are curtailed by a uniform complexity bound on EFTs. To draw a precise connection, we stress that this requires more than purely referring to potentially infinite discrete sets of manifolds.  Since all of the Calabi-Yau manifolds are believed to be connected via flop and conifold transitions \cite{reid1987moduli}, there might even exist a single theory that allows one to explore all Calabi-Yau vacua simultaneously \cite{Sen:2025bmj,Sen:2025oeq}. If such an idea can be made precise and one has infinitely many topologically distinct compact Calabi-Yau manifolds, we expect to find a contradiction with the Finite Complexity Conjecture.  Intuitively, this stems from the infinitely many connections in the field space that appear in conflict with having a finite cell decomposition. Our perspective using EFT domains fits with the expectation that there are indeed finitely many compact Calabi-Yau threefolds that might be connected by transitions. In this case we can consider the four-dimensional EFTs arising from compactifications of Type II string theory that are expansions around a Minkowski vacuum. Depending on the cutoff, we find EFTs on individual domains, as depicted schematically for the EFTs near a conifold transition in figure \ref{fig: conifold transition}. The moduli space $\cM_{\rm QG}= \lim_{\Lambda \rightarrow 0} \cM_{{\rm QG};\Lambda}$ is then naturally identified with the irreducible moduli space proposed by Reid \cite{reid1987moduli}, and  has finite complexity. This will be contrasted with Anti-de Sitter (AdS) compactifications momentarily.   

Before turning to AdS compactifications, let us give a related illustrative example how the Finite Complexity Conjecture restricts discrete infinities, consider four-dimensional 
$\text{SU}(N)$ Yang–Mills theory with $\cN=2$ supersymmetry. While for small values of 
$N$ such theories are expected to admit embeddings into string theory (see e.g.~\cite{Kumar:2010ru,Morrison:2011mb,Kim:2019ths,Tarazi:2021duw,Martucci:2022krl}), the Finite Complexity Conjecture implies the existence of an upper bound on $N$. Indeed, as $N\rightarrow \infty$, not only does the number of light fields increase, but the complexity of the associated Coulomb branch EFTs grows as well \cite{Carrascal:2025vsc}.

\paragraph{AdS compactifications.} Let us now turn to compactifications of string theory and M-theory that lead to EFTs only valid around an AdS vacuum. These compactifications are on backgrounds of the form AdS$_d \times K$, where $K$ is positively curved and supported by background fluxes. Famously, there are infinite series of such vacuum solutions within string theory, as demonstrated for example by the AdS$_5\times S^5$ compactifications of Type IIB string theory with $N$ units of $F_5$ form flux through the $S^5$. Denoting by $M_{\rm pl}$ the five-dimensional Planck mass, we note that the effective theory will admit an unavoidable cutoff $\Lambda_{\rm t}\propto M_{\rm pl} N^{-2/3}$, since at this scale Kaluza-Klein modes of $S^5$ will become relevant. Requiring that $\Lambda_{\rm t}>\Lambda$, for a fixed cutoff $\Lambda$, we find that only finitely many values of $N$ contribute to the EFT landscape parameterized by $\cM_{{\rm QG};\Lambda}$. Recalling our definition of $\cM_{\rm QG} = \lim_{\Lambda \rightarrow 0} \cM_{{\rm QG};\Lambda}$, we recognize that this would imply that $\cM_{\rm QG}$ contains a discrete infinite set of EFTs. Using the AdS/CFT duality it was argued that these theories are genuinely different EFTs \cite{Banks:2025nfe,Sen:2025oeq}, which sheds doubt on claiming the tameness of $\cM_{\rm QG}$. This does, however, not contradict our conjecture about the individual $\cM_{{\rm QG};\Lambda}$, but indicates that in the AdS case, tameness requires additional bounds. This is consistent with the conjectures of \cite{Douglas:2023fcg} for the tameness of the space of CFTs and the recent discussion in \cite{Baykara:2025nnc}. In fact, tameness and complexity turn out to be a powerful tool to analyze and `count' AdS vacua as we discuss in the next subsection. In addition, we will also speculate if $\cM_{\rm QG}$ is of finite complexity by an \textit{emergence of simplicity} argument in the spirit of the mathematical toy models presented at the end of section~\ref{sharp-o-complexity}.

\subsection{Counting EFTs, volume growth, and complexity}
\label{sec:complexity_volume_counting}

In this section we discuss how the Finite Complexity Conjecture relates to other recently stated swampland constraints, such as volume growth and the counting of vacua. We then suggest how these results can be used to turn the abstract bounds claimed by the conjecture into explicit numbers. Note that we will only focus on the complexity of the spaces $\cM_{{\rm QG};\Lambda}$ and $\cM_{\rm QG} = \lim_{\Lambda \rightarrow 0} \cM_{{\rm QG};\Lambda}$ and leave a study of the complexities of the EFT Lagrangians to future work.

\paragraph{Distance functions on $\cM_{\rm QG}$.} To apply the Finite Complexity Conjecture we have to gain a better handle on $\cM_{\rm QG}$ and $\cM_{{\rm QG};\Lambda}$. These spaces are rather abstractly defined and have a known structure only in specific sectors. For example, we have recalled in section~\ref{complexity_comp} that $\cM_{\rm QG}$ can contain the moduli spaces of complex structure deformations of Calabi-Yau manifolds. These components are Riemannian manifolds with a well-defined physical metric, the Weil-Petersson metric. In contrast, $\cM_{\rm QG}$ might contain discrete points associated to a flux lattice or many disconnected components of varying dimension. This happens, for example, in the AdS compactifications discussed in section~\ref{complexity_comp}. In these general cases there is no naturally associated notion of distance that indicates how close two disconnected pieces of $\cM_{\rm QG}$ are to each other. Candidate distance functions on such disconnected or non-Riemannian spaces have already been discussed in \cite{Douglas:2010ic} and recently in \cite{Li:2023gtt,Shiu:2023bay,Basile:2023rvm,Mohseni:2024njl,Debusschere:2024rmi,Palti:2025ydz}. While a complete picture has yet to be developed, these constructions motivate the expectation that $\cM_{\rm QG}$ admits a natural description as a metric space with a well-defined distance function.

\paragraph{Counting of EFTs.} Let us now assume that the Finite Complexity Conjecture is satisfied and consider the space $\cM_{{\rm QG};\Lambda}$. We can then attempt to count the number of distinct effective theories existing on this space. While this question is easily posed, it is subtle to decide how to perform the counting. One natural way to proceed is to consider the finite EFT covering  $\mathcal{U}=\{ U_i\}_{i=1,...,n_{\mathcal{U}}}$ as defined at the end of section~\ref{sec:conjecture}. A natural way to define the number of EFTs is the number of domains $n_{\mathcal{U}}$, which corresponds to the number of EFT Lagrangians needed to cover $\cM_{\rm QG;\Lambda}$. To get a universal result, we can take the infimum over all coverings and define 
\beq \label{number-EFTs}
n(\cM_{{\rm QG};\Lambda}) = \text{inf}_{\cU} (n_{\cU})\ .
\eeq 
As stressed above the Finite Complexity Conjecture for $\cM_{\rm QG;\Lambda}$ ensures that $n(\cM_{{\rm QG};\Lambda})$ is \textit{finite}.
While this counting takes into account the validity of the EFTs, via our definition of $\cM_{{\rm QG};\Lambda}$ and the notion of an EFT covering, it does not account for the fact that some EFTs might have larger field spaces and hence can account for wider ranges of couplings. 

\paragraph{Weighted counting of EFTs.}
It is a long-standing challenge to suggest a physically plausible and mathematically well-defined weight-factor modifying \eqref{number-EFTs}. An interesting suggestion was recently put forward \cite{Baykara:2025nnc}, where it was argued a natural choice of weight-factor is to account for the volume of the field space in \textit{flat and tachyonic directions}. Assuming it to be physically well-motivated, one faces various challenges to make such a definition precise. It turns out that the Finite Complexity Conjecture ensures that such a weight-assignment can be made precise. Firstly, while volume is a well-defined notion for a Riemannian manifold, we have to specify what takes the role of the volume on a general metric space. The Finite Complexity Conjecture ensures that the so-called Hausdorff measure, defined on a metric space by covering the space with metric balls and minimizing over all possible covers, becomes a well-defined volume measure. In fact, if the considered metric space is tame, it assigns finite volumes Vol$(A)$ to bounded sets (see \cite{Barroero2014} for details). 

Secondly, for a generic potential it can be difficult to distinguish flat and tachyonic directions without a clear handle on the structure of the potential, and it generally does not lead to finitely many regions. Once again the Finite Complexity Conjecture, implying the sharp o-minimality of the potential, can resolve this situation. Firstly, there are finitely many regions of the field space on which $\partial V = 0$ \cite{VdDries} and finitely many regions with Hessians of fixed rank and definite signature. Secondly, all such regions are tame sets and the number of such regions can be constrained by the complexity of the Lagrangian $(\cF_{\rm EFT},\cD_{\rm EFT})$, which is universally bounded by $(\mathfrak{F}_\Lambda,\mathfrak{D}_\Lambda)$.
For simplicity we will consider only flat directions in the following, leaving the more general discussion to future work. 

\paragraph{Counting with flat potential.} Let us now consider the space of EFTs that are valid at least up to $\Lambda$ and have a vanishing potential, whose moduli space we denote by $\cM_{\text{flat};\Lambda}$. We introduce a EFT cover $\cU = \{U_i\}_{i=1,...,n_\cU}$, for this space, with the properties outlined at the end of section~\ref{sec:conjecture}. We now replace the direct count \eqref{number-EFTs} by 
\beq \label{counting1}
   \mathfrak{m}_{\mathcal{U}}(\cM_{\Lambda}) = \sum_{i=1}^{n_{\cU}} \text{Vol}_{d_i}(U_i)\ . 
\eeq
This counts flat directions with a volume factor as suggested in \cite{Baykara:2025nnc} and is well-defined and finite due to the Finite Complexity Conjecture. In particular, it is crucial to have a fixed dimension $d_i$ of $U_i$, a finite covering $\cU$, and assigned Vol$_{0}(U_i) = |U_{i}|$ to zero-dimensional sets. To remove the dependence on $\cU$, we again take the infimum, but this time picking the $\cU$ that minimizes the volume measure:
\beq \label{frak-n}
 \mathfrak{n}(\cM_{\text{flat};\Lambda})
 =\inf_{\mathcal{U}}\sum_i \text{Vol}_{d_i}(U_i)\ .
\eeq
Since the Hausdorff measure agrees with the standard volume in Riemannian manifolds, we expect that \eqref{frak-n} gives a formal generalization of the counting function introduced in \cite{Baykara:2025nnc}. In the following, we note on how the expression \eqref{frak-n} relates to complexity and comment on the expected scaling with $\Lambda$ connecting with \cite{Baykara:2025nnc}. 

\paragraph{Complexity and embeddings.} In order to evaluate \eqref{frak-n} in terms of complexity, and to study the Finite Complexity Conjecture itself, we face the challenge to assign a complexity to the spaces $\cM_{\text{flat};\Lambda}$ and $\cM_{\text{flat}} = \lim_{\Lambda\to 0}\cM_{\text{flat};\Lambda}$, a process which should account for their metric structure. Since a complexity theory for metric spaces and Riemannian manifolds has not been developed so far, one way to nevertheless use the complexity theory of section \ref{sec:tameness} is trying to embed $\cM_{\text{flat};\Lambda}$ and $\cM_{\text{flat}}$ into some sufficiently large $\bbR^N$ while preserving the metric structure. In particular, this embedding should be done isometrically for all components of $\cM_{\text{flat}}$ that are Riemannian manifolds, i.e.~such that the Euclidean metric of $\bbR^N$ reduces to this metric. Upon embedding $\cM_{\text{flat};\Lambda}$ and $\cM_{\text{flat}}$ into $\bbR^N$, there is a natural way to assign a complexity to these spaces if the image is definable in a sharply o-minimal structure. As explained in \cite{Grimm:2025lip}, the tameness of the embedding precisely reproduces the compactifiability criterion proposed in \cite{Delgado:2024skw}, which states that the volume of a geodesic ball in moduli space grows asymptotically no faster than in Euclidean space. While we eventually hope to avoid a detour via such isometric embeddings, we will see that in case one uses the available results for these cases, one can make the relation between volume and complexity explicit. In fact, the relation between these two quantities is a prominent part of tame geometry.

\begin{figure}[h!]
    \centering
    \includegraphics[width=0.35\linewidth]{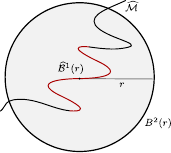}
    \caption{A segment $\widehat{\mathcal{B}}^1(r)$  of length $2r$ (in red) of the embedded curve $\widehat{\mathcal{M}}$ inside an Euclidean disk of radius $r$, corresponding to $\kappa=1$ and $N=2$. The curve is bent strongly and occupies a subdisk of smaller radius than $\widehat{\cM}_r= \widehat{\cM}\cap B^2(r)$ ensuring the length of $\widehat{\mathcal{B}}^1(r)$ is smaller than the length of $\widehat{\cM}_r$.}
    \label{fig:volumeslice}
\end{figure}

\paragraph{Volume growth and complexity.} Let us first consider a component $\cM \subset \cM_{\text{flat}}$ which is a $\kappa$-dimensional Riemannian manifold with metric $g$ and therefore admits a isometric embedding $\widehat \cM\subset \bbR^N$ for some $N$ by the Nash embedding theorem. We now assume that $\widehat \cM$ is definable in a sharply o-minimal structure in order that the Finite Complexity Conjecture holds. Considering a Euclidean $N$-dimensional ball 
$B^{N}(r)$ around some base-point $x_0 \in \widehat \cM$ and of radius $r$ in $\bbR^N$, we define $\widehat \cM_r =  B^N(r) \cap \widehat \cM$. As shown in \cite{yomdin2004tame}, the volume of $\widehat \cM_r$ 
satisfies the inequality
\begin{equation} \label{volumegrowth}
\text{Vol}_\kappa(\widehat \cM_r) \leq c(\kappa,N) \, \cC(\cF,\cD)\,   r^\kappa\,.
\end{equation}
Here $\cC(\cF,\cD)$ is a coefficient depending on the complexity of the embedding of $\cM$, and  
\begin{equation} \label{c-coeff}
    c(\kappa,N) = \text{Vol}_\kappa(B^\kappa(1))\,  \frac{\Gamma \big(\frac{1}{2} \big)\Gamma \big(\tfrac{N+1}{2} \big) }{\Gamma \big( \tfrac{ \kappa+1}{2} \big)\,\Gamma \big( \tfrac{N-\kappa-1}{2} \big)} \ , 
\end{equation}
where $\text{Vol}_\kappa(B^\kappa(1))$ is the volume of the $\kappa$-dimensional ball of radius 1. The condition \eqref{volumegrowth} has multiple consequences. In particular, we find that the volume of any $\kappa$-dimensional geodesic ball $\mathcal{B}^\kappa (r) \subset \cM$ of radius $r$ grows maximally as fast as Euclidean space \cite{Grimm:2025lip} reproducing the quantum gravity criterion of \cite{Delgado:2024skw}. This follows from the geometric fact that, after choosing matching center points, the embedding of $\mathcal{B}^\kappa (r)$ is a subset of $\widehat{\mathcal{B}}^\kappa (r) \subset \widehat \cM_r$ such that 
\beq
\text{Vol}_\kappa (\widehat{\mathcal{B}}^\kappa (r)) \leq \text{Vol}_\kappa(\widehat \cM_r)\ , 
\eeq 

as depicted in figure~\ref{fig:volumeslice}. Note that if $\cM$ has high curvature, then the embedding of $\mathcal{B}^\kappa(r)$ may only occupy a comparably small subregion of $\widehat \cM_r$. Since $\text{Vol}_\kappa (\mathcal{B}^\kappa (r))=\text{Vol}_\kappa (\widehat{\mathcal{B}}^\kappa (r))$, we thus conclude that
\beq \label{volumegrowth2}
   \text{Vol}_\kappa(\mathcal{B}^\kappa (r)) \leq c(\kappa,N) \, \cC(\cF,\cD)\,   r^\kappa\ .
\eeq
Note, however, that \eqref{volumegrowth2} is stronger than the proposal in \cite{Delgado:2024skw} and holds not only asymptotically in $r$, but for every geodesic ball $\mathcal{B}^ \kappa(r) \subset \cM$ at every point of $\cM$ since $\kappa,N$ and $(\cF,\cD)$ are global data associated to $(\cM,g)$ and its embedding. In fact, the prefactor $c(\kappa,N)$ can also be upper bounded by a function of $\cF$ and $\cD$, since we know that $\kappa \leq \cF$ and $N \leq \cO(\cF^{3})$.\footnote{Note that $N \leq \cO(\cF^{3})$ is a weak bound that is inferred using the assertion that there is a $\cC^k$ embedding. More precisely, the $\cC^k$ Nash embedding theorem gives for non-compact Riemannian $\cM$ an upper bound $N\leq \frac{1}{2}\kappa(\kappa + 1)(3\kappa + 11)$.} This implies that there is always a bound of the form 
\beq \label{volumegrowth3}
  \text{Vol}_\kappa(\mathcal{B}^\kappa(r)) \leq \mathfrak{C}(\cF,\cD)\,   r^\kappa\,
\eeq
in which the prefactor no longer has an explicit dimensional dependence and is only a function of the complexity $(\cF,\cD)$. Until there is a developed theory of complexity of Riemannian manifolds, we can deduce the form of $\mathfrak{C}(\cF,\cD)$ via the embedding. Conversely, we can infer for a given $(\cM,g)$ the coefficient $\mathfrak{C}(\cF,\cD)$ by investigating the volume of all geodesic balls $\mathcal{B}^\kappa(r)$ in $\cM$. In the later case, we are guaranteed that such a universal $\mathfrak{C}(\cF,\cD)$ exists when demanding the existence of a sharply o-minimal isometric embedding.

\paragraph{Complexity dependence.} The volume bounds \eqref{volumegrowth2}, \eqref{volumegrowth3} provide a natural relation between the counting \eqref{frak-n} and the complexity of $\cM_{\text{flat};\Lambda}$. Note, however, that we cannot display a universal formula for $\cC(\cF,\cD)$ or $\mathfrak{C}(\cF,\cD)$. While the functional form of $\mathfrak{C}$ and $\cC$ will not depend on the detailed geometry of $\cM_{\text{flat};\Lambda}$, they will depend on the underlying sharply o-minimal structure. One expects that universal formulas can be found relating these coefficients to the counting function $P_\cF(\cD)$ associated to the sharply o-minimal structure under consideration (see axiom (iv) in section~\ref{sharp-o-complexity}). One can show, see \cite{binyamini2022sharply} for details, that $\cC(\cF,\cD)$ is \textit{always} a polynomial in $\cD$, but \textit{generically} depends exponentially on $\cF$. We write this as 
\beq 
    \cC, \mathfrak{C} \sim \text{poly}_\cF(\cD)\ ,
\eeq
where $\text{poly}_\cF$ represents a generic polynomial function in $\cD$ (distinct for $\cC$ and $\mathfrak{C}$).

There is, however, an interesting way to circumvent an exponential growth in $\cF$: if one examines the dimensional coefficient $c(n,N)$, given in \eqref{c-coeff}, one realizes that it depends on the $\kappa$-ball volume $ B^\kappa(1)$ which super-exponentially decays as $\kappa^{-\kappa/2}$ in the limit $\kappa \rightarrow \infty$.  This fast fall-off might be compensated by the growth of $\cC$ or the $\Gamma$-prefactors in \eqref{c-coeff}, but it is not hard to give examples for which the $\mathfrak{C}(\cF,\cD)$ inherits this rapid fall-off. Consider a component $\cM$ given as a degree $\cD$ polynomial hypersurface in $\bbR^{\cF}$ with the induced metric, we find that $\cC(\cF,\cD) \sim \cD$ (see appendix~\ref{app_volumebounds} for details). For example, this equation might specify a $\cF-1$-dimensional sphere embedded into $\bbR^{\cF}$, where one has $\cD=2$. In the hypersurface case, since $N=\cF$ and $\kappa=\cF-1$, we have $c(\kappa,N) \sim \cF^{-\cF/2}$ for large $\cF$ such that
\beq \label{suppressed_C}
   \mathfrak{C}(\cF,\cD) \sim \cF^{-\cF/2} \cD\ . 
\eeq
This super-exponential suppression was key in the arguments of \cite{Baykara:2025nnc} to argue for a specific cutoff dependence of $\mathfrak{n}(\cM_{\text{flat};\Lambda})$ in the limit $\Lambda \rightarrow 0$. Let us note that this strong suppression cannot be universally inferred in our setting. For example, in order that $c(\kappa,N)$ gives a strong suppression, one needs that the embedding dimension grows as $N \sim \kappa$ and not with a higher power of $\kappa$. While Nash's result gives an upper bound $\cO(\kappa^3)$, it is conceivable that tameness constrains this $N$ drastically. Furthermore, as shown in appendix~\ref{app_volumebounds}, also the growth of $\cC(\cF,\cD)$ with $\cF$ can be exponential. Since in such limits also our upper bounds, like Vol$_\kappa (\mathcal{B}^\kappa (r)) \leq \text{Vol}_\kappa(\widehat \cM_r) $, will get worse, we believe that there must be a cleaner argument for the suppression of the type \eqref{suppressed_C} when avoiding the embedding and rather using~\eqref{volumegrowth3}. We leave this as an interesting mathematical task for future work.

\paragraph{Cutoff dependence and distances.}
Let us close this section by combining the volume growth \eqref{volumegrowth3} and the weighted counting of vacua \eqref{frak-n}. In order to do that we consider a finite the EFT covering $\cM_{\text{flat};\Lambda}=\bigcup_i U_i$, where we recall that we focus on EFTs with vanishing potential. The full moduli space will be denoted $\cM_\text{flat} = \lim_{\Lambda \rightarrow 0} \cM_{\text{flat};\Lambda} $.
We take the maximal geodesic distance that can be traversed in each $U_i$ to be $2r_i$ and place each $U_i$ into a $d_i$-dimensional geodesic ball $\mathcal{B}^{d_i}_i(r_i)$, i.e.~we have $U_i \subset \mathcal{B}^{d_i}_i(r_i) \subset\cM_{\text{flat}}$. We then refer to the complexity of the ball $\mathcal{B}^{d_i}_i(r_i)$ by $(\cF_i,\cD_i)$, which yields the upper bound 
\beq
   \text{Vol}_{d_i}(U_i)\leq  \text{Vol}(\mathcal{B}^{d_i}_i(r_{i}))  \leq \mathfrak{C}(\cF_i,\cD_i)\,   |r_i|^{d_i}\ . 
\eeq
Note that the Distance Conjecture implies that $r_{\rm max} \sim \frac{1}{\alpha}|\log \Lambda|$, where $\alpha$ is an order-one constant that we will suppress in the following. Inserted into \eqref{frak-n}, we find 
\beq \label{nM-explicit}
  \mathfrak{n}(\cM_{\text{flat};\Lambda}) \lesssim \inf_{\mathcal{U}}\sum_{i=1}^{n_\cU} \mathfrak{C}(\cF_i , \cD_i) |\log \Lambda|^{d_i} \ . 
\eeq
Given that $\cM_{\text{flat};\Lambda} \subset \bigcup_i \mathcal{B}^{d_i}_i(r_{i})$, we now recall that by the axioms of sharp o-minimality we have $\cF_\Lambda \leq\max(\cF_i) $ and $\cD_\Lambda \leq \sum_i\cD_i$ and where $(\cF_\Lambda,\cD_\Lambda)$ is the complexity of $\cM_{\text{flat};\Lambda}$.\footnote{Here one needs to be careful and recall that the axioms are statements about the FD-filtration. It might still be possible that one can find a more optimal representation of the whole space $\cM_{\text{flat};\Lambda}$ with a smaller $\cF$ or $\cD$.}
Assuming that $\cM_\text{flat}$ is a moduli space of finite complexity $(\cF,\cD)$, we note that the expression \eqref{nM-explicit} is a polynomial in $|\log \Lambda|$ with cutoff independent coefficients,  recovering the claim of \cite{Baykara:2025nnc} that $\mathfrak{n}(\cM_{\text{flat};\Lambda}) \lesssim a |\log \Lambda|^b $, where $a,b$ are complexity-dependent coefficients. 

As mentioned in section~\ref{complexity_comp} the existence of a sharply o-minimal $\cM_\text{flat}$ is plausible for the present case of compactifications of string theory without potential. In case that $(\cF_{\Lambda},\cD_{\Lambda})$ diverges in the limit $\Lambda \rightarrow 0$ limit, like it is the case for many AdS vacua as discussed in section~\ref{complexity_comp}, one needs to refine and extend the argument. A plausible extension takes into account that in this case one should account for the strong suppressions such as the one shown in \eqref{suppressed_C} as $\cF_i \rightarrow \infty$ and claim that small values of $\cF_i$ dominate the sum. It would be desirable to extend the discussion to such cases in the future. We suspect that the Finite Complexity Conjecture will give sufficient controls over these limits as well.

\section{Conclusion and outlook}\label{sec:conclusion}

In this work we used sharp o-minimality to develop a quantitative framework for capturing the complexity of effective field theories (EFTs) in the quantum gravity landscape. A central claim is that the seemingly infinite Wilsonian data of an EFT can be encoded in a finite amount of information, often through implicit structures such as differential constraints, yielding a description of finite tame complexity. Motivated by swampland finiteness arguments, we formulated a Finite Complexity Conjecture. In its local form, the conjecture asserts that every quantum-gravitational EFT admits a description of finite complexity $(\cF_{\rm EFT},\cD_{\rm EFT})$. In its stronger global form, it states that for EFTs valid at least up to a fixed cutoff $\Lambda$, these complexities are uniformly bounded across the set of EFTs $\cM_{\rm QG;\Lambda}$. Furthermore, we proposed that $\cM_{\rm QG;\Lambda}$ itself has finite complexity $(\cF_\Lambda,\cD_\Lambda)$. A key ingredient in making these statements precise is the notion of EFT domains and EFT coverings, which capture the intrinsically local nature of EFT descriptions over parameter and moduli spaces and provide a more rigorous way to describe the space of EFTs with a fixed cutoff. This perspective also leads to well-defined notions of counting and volume measures, whose finiteness is naturally enforced by the global conjecture. We supported the conjecture by connecting it with other swampland principles and by working through a set of illustrative string-theoretic examples.

Beyond providing a rigorous mathematical foundation, the Finite Complexity Conjecture has concrete consequences for how finiteness manifests in low-energy physics. For individual EFTs, finite tame complexity constrains not only the light spectrum but also the complexity of allowed interactions and scalar potentials. For families of theories, tameness enforces finiteness of EFT domains and allows for defining sharpened notions of counting: algebraic or differential constraints on EFT data, in particular on the scalar potential, can be used to define  restrictions of the EFT domains, and the resulting smaller sets will inherit finite tame complexity. This places volume-weighted counting prescriptions, such as those proposed in \cite{Baykara:2025nnc}, on a solid footing. Moreover, as discussed in section~\ref{sec:complexity_volume_counting}, tame complexity bounds the volume of an embedded space, providing a measure that is neither purely topological nor sensitive to detailed numerical data. Even in simple cases, such as the moduli space of ten-dimensional Type IIB string theory, tameness relies crucially on duality identifications: without them no tame embedding exists \cite{Grimm:2025lip}, and the compactifiability criterion of \cite{Delgado:2024skw} is violated. The same structural role of dualities appears on the EFT side, where they are needed to maintain a finite tame complexity of the Lagrangian descriptions, as discussed for $\cN=2$ super-Yang-Mills theory in \cite{Carrascal:2025vsc}. The example of type IIB in 10 dimensions, as well as the other cases discussed in section \ref{complexity_comp}, provide evidence that quotienting by the duality action is required to achieve descriptions of finite complexity. In light of the finite complexity conjecture,  a divergence in the evaluation of the complexity of an effective theory expected to be compatible with quantum gravity suggests the existence of additional duality relations which have not been properly considered.
%

An important challenge for the future is to make the bounds in the Finite Complexity Conjecture explicit, i.e.~to determine the $\Lambda$-dependence of $(\cF_\Lambda,\cD_\Lambda)$ and $(\mathfrak{F}_{\Lambda},\mathfrak{D}_{\Lambda})$. It is tempting to expect at most polynomial and/or logarithmic growth as $\Lambda\rightarrow 0$ (in Planck units), but deriving such quantitative scalings appears difficult at present. Progress may nevertheless be possible in restricted sectors of the landscape, for instance within specific classes of string compactifications or in settings with high amounts of supersymmetry, where additional structure yields much sharper finiteness statements. As one illustration, remarkably strong and explicit bounds on the massless matter spectrum of six-dimensional $\cN=(1,0)$ supergravity theories have recently been obtained in \cite{Kim:2024eoa,Birkar:2025gvs,Birkar:2025rcg}.

The Finite Complexity Conjecture also strengthens the case for further developing tame geometry and for making its results quantitative using sharp o-minimality. At the same time, it highlights where such developments are likely to be most impactful for physics and can provide quantitative results for quantum field theory and string theory. Physics examples indicate that both differential constraints and recursion relations are the basic mechanism ensuring finite tame complexity. In particular, periods and their variation over parameter spaces are ubiquitous in string theory and often govern effective couplings and potentials. Establishing a firm notion of tame complexity for periods is a natural and important next step, which would follow from a proof of the sharp o-minimality of the structure $\bbR_{\rm LN}$. Beyond this, it will be important to extend the tame complexity methods to settings where EFT information is only accessible asymptotically, for instance through multi-summable transseries, since resurgent and multi-summable behavior is a common feature of perturbative expansions in QFT and string theory.
Finally, our analysis underscores the need for an intrinsic complexity theory of Riemannian manifolds and metric spaces, independent of isometric embeddings. Developing this would be a major advance, enabling embedding-independent control of volume growth and counting in the landscape.

A major extension of our proposal is to formulate finiteness of tame complexity without the restriction to EFTs or a fixed space-time dimension. This could, in principle, allow one to move beyond the cutoff-restricted landscape $\cM_{\rm QG;\Lambda}\subseteq \cM_{\rm QG}$, at the price of systematically incorporating dual descriptions that may change the effective dimensionality or even require non-EFT descriptions.
An illustrative example is a decompactification limit in moduli space, where an infinite tower of Kaluza-Klein modes becomes light, the lower-dimensional EFT ceases to be valid and the apparent complexity diverges.
Yet the same physics is now captured by a higher-dimensional EFT with finitely many fields,   providing an effective description with finite tame complexity.   
Similarly, in other asymptotic regimes where weakly coupled asymptotically tensionless strings are expected to emerge \cite{Lee:2019wij}, the worldsheet theory may furnish the finite-complexity description underlying an infinite tower of string excitations.
This realization of \textit{emergence of simplicity} and incorporation of dual descriptions is natural in the framework of sharp o-minimality, as indicated in section~\ref{sharp-o-complexity}. Other potential directions in which our proposal may be extended are the definition of complexity of field theories beyond two derivatives and beyond Lagrangian theories, in order to verify whether the consistency with quantum gravity allows for a generalized formulation of the finite complexity conjecture in these settings. This will require improvements in our understanding of resurgence and tameness, as well as to study the sharp o-minimality of alternative objects which provide a systematic characterization of the complexity of a theory.  More broadly, it is tempting to view finiteness of tame complexity as a basic structural constraint on physical description: nature only ever provides finitely many observable data at finite resolution, and whenever an infinite amount of information seems required, it may signal that we are using a wrong framing rather than revealing genuine physical complexity.

\subsubsection*{Acknowledgements}
We would like to thank Kaan Baykara, Gal Binyamini, Gonzalo F. Casas, Alberto Castellano, Martín Carrascal, Raf Cluckers, Ferdy Ellen, Bruno Klingler, Amineh Mohseni, Jeroen Monnee, Fabian Ruehle, Cumrun Vafa, Stefan Vandoren, and David Wu for valuable discussions and comments. Furthermore, we are grateful to the CMSA and the Swampland Initiative for their financial support and hospitality during the completion of this work. This research is supported, in part, by the Dutch Research Council (NWO) via a Vici grant.


\appendix

\section{Volume bounds for semi-algebraic and Pfaffian sets} \label{app_volumebounds}

Tame geometry imposes strong bounds on the volume growth of its definable objects, which closely mirror our intuition about volumes in Euclidean space. The main result in this context was already presented and used in section \ref{sec:complexity_volume_counting}. We will now explain it in more detail. 

For any tame set $A\subset \mathbb{R}^N$ and  for any $0\leq \kappa\leq N$ there exists an integer number $b_{0,N-\kappa}$ such that for any $(N-\kappa)$-dimensional affine plane, $P$, the number of connected components of $A\cap P$ is bounded by $b_{0,N-\kappa}$  \cite{yomdin2004tame}. Let now $\kappa$ be the dimension of $A$, then for any $N$-dimensional ball $B^N(r)$ in $\mathbb{R}^N$, 
        \begin{equation}
            {\rm Vol}_\kappa(A\cap B^N(r))\leq  c(N,\kappa)\, b_{0,N-\kappa}(A)\cdot r^\kappa\,,
            \label{eq: Yomdin bound}
        \end{equation}
        where and $c(n,l)$ is a normalization constant given by
        \begin{equation}
         c(N,\kappa)={\rm Vol}_\kappa(B^\kappa(1))\cdot\frac{\Gamma\left(\frac{1}{2}\right)\Gamma\left(\frac{N+1}{2}\right)}{\Gamma\left(\frac{\kappa+1}{2}\right)\Gamma\left(\frac{N-\kappa+1}{2}\right)}\,,
        \end{equation}
        with ${\rm Vol}_\kappa (B^\kappa(1))$ the Euclidean volume of the $l$-dimensional ball of radius one.

For generic tame sets, estimating the bound $b_{0,N-\kappa}(A)$ is extremely challenging. However, in the cases in which $A$ is not only tame, but can be defined in a sharply o-minimal structure we can obtain a much greater control over the bounds. Indeed, one can relate $b_{0,N-\kappa}(A)$ with the format and degree of $A$ and use the number of connected components of the intersection with hyperplanes as a proxy for the complexity of the set. In other words, there exists a function $\mathcal{C}(\cF,\cD)$ such that if $A$ is definable in a sharply o-minimal structure with $A\in \Omega_{F,D}$ then $b_{0,N-\kappa}(A)\leq \mathcal{C}(\cF,\cD)$.  

Consider, for example, the case when $A$ is a semi-algebraic set defined by polynomial equations in $\bbR^N$: 
\beq
    A= \bigcup_{i=1}^K \{P_i(x) = 0\ \text{or} \ P_i(x)>0\}\ . 
\eeq
Denoting by $d_i$ the degree of $P_i$ one finds \cite{yomdin2004tame}
\beq
  b_{0,N-\kappa}(A) \sim \sum_{i=1}^K d_i^{N-\kappa}\,,
\eeq
and we can establish the relation with the format and the degree of $A$ explicitly using the axioms of sharp o-minimal structures presented in section \ref{sharp-o-complexity}. First, the format in semialgebraic sets is given by the number of variables, so $\cF=N$. Second, the degree is obtained by summing the degree of all the polynomials involved in the definition of $A$, and thus $\cD= \sum d_i$. Therefore we can take the bound
\begin{equation}
    b_{0,N-\kappa}(A) \leq \sum_{i=1}^K \cD^{\cF} = \mathcal{C}(\cF,\cD)\,,
    \label{eq: gavrielov bound algebraic}
\end{equation}
and verify that $\mathcal{C}(\cF,\cD)$ grows polynomially with the degree of the set $A$ and exponentially with its format. 

Note that in the particular case in which $A$ is a hypersurface (and thus $\kappa=1$), we have a much better bound
\begin{equation}
    b_{0,N-\kappa}(A) \leq \cD\,,
\end{equation}
which showcases that the generic bounds of the form \eqref{eq: gavrielov bound algebraic} obtained using sharp o-minimal principles are far from optimal. This is to be expected, as the example under consideration is particularly simple (semi-algebraic) and thus possesses a richer structure that allows to make more precise estimations. Nevertheless, the present computation helps to illustrate how the formalism works and gives valuable intuition about its potential application to more general settings, where sharper bounds might not be available. 

For the semi-Pfaffian case, the relations are more involved, but we can still provide explicit bounds. For instance, given a $\kappa$-dimensional variety defined as the zero locus of a family of Pfaffian functions $A=\{x\in U| f_1=\dots=f_{N-\kappa}=0\}$, where $f_i$ are defined in a domain $U\subset \mathbb{R}^N$ and have a common Pfaffian chain of order $r$ and degrees $(\alpha,\beta_i)$ respectively, then the number of connected components of the set $A\cap P$, with P a $(n-l)$-dimensional hyperplane, is bounded by \cite{Khovanskii, Fewnomials, GabVor04} 
\begin{equation}
b_{0,N-\kappa}(A)\leq 2^{r(r-1)/2+1}\beta(\alpha+2\beta-1)^{n-1}((2n-1)(\alpha+\beta)-2n+2)^r\,,
\end{equation}
where $\beta:=\max_{1\leq i\leq\kappa} \{\beta_i\}$. Notably, this bound has no explicit dependence on the dimension of the variety, only on the embedding space. Taking the format and the degree of the Pfaffian variety  to be $\mathcal{F}=r+n$ and $\mathcal{D}=\alpha+\beta$, one can find again a coarser bound of the form
\begin{equation}
b_{0,N-\kappa}(A)\leq {\rm poly}_\mathcal{F
}(\mathcal{D
})\,.
\end{equation}

\bibliography{literature}
\bibliographystyle{utphys}

\end{document}